\documentclass[longauth]{aa}  

\usepackage{graphicx}
\usepackage{float}
\usepackage{amsmath}
\usepackage{txfonts}
\begin{document} 

\def\simgt{\lower.5ex\hbox{$\; \buildrel > \over \sim \;$}}
\def\simlt{\lower.5ex\hbox{$\; \buildrel < \over \sim \;$}}
\def\etal{{\it et al.}}
\def\msun{M$_\odot$}
\def\rsun{R$_\odot$}
\def\Teff{$T_{\rm eff}$}
\def\teff{$T_{\rm eff}$}
\def\BMV{$B-V$~}
\def\EBV{$E_{B-V}$}
\def\dv{$\Delta V$~}
\def\Mc{$M_\rm c$~}
\def\X{$X$~}
\def\FeH{${\rm [Fe/H]}$~}
\def\feh{${\rm [Fe/H]}$~}
\def\Y{$Y$~}
\def\Z{$Z$~}
\def\Lto{$L_{\rm TO}$~}
\def\Hp{$H_{\rm p}$~}
\def\logL{$\log (L/L_{\odot})$~}
\def\Mv{$M_{\rm v}$~}
\def\M{$M$}
\def\R{$R$}
\def\alfamlt{$\alpha_{\rm MLT}$~}
\def\alfaatm{$\alpha_{\rm atm}$~}
\def\alfa2d{MLT--$\,\alpha^{\rm 2D}$~}
\def\afa{$\alpha$~}
\def\taufot{$\tau_{\rm ph}$~}
\def\logg{$\log (g)$}
\def\Dnu{$\Delta\nu$~}
\def\dnu{$\Delta\nu$~}
\def\deltanu{$\Delta\nu$~}
\def\numax{$\nu_{\rm max}$}
\def\numaxm{\nu_{\rm max}~}
\def\Dnum{\Delta_{\rm nu}~}
\def\vmic{v_{\rm mic}~}

\def\kepler{\mbox{\textit{Kepler}}~}
\def\Space{\mbox{SP\_Ace}~}

   \title{RAVE stars in K2}

   \subtitle{I. Improving RAVE red giants spectroscopy using asteroseismology from K2 Campaign 1}

   \author{M.~Valentini
          \inst{1}
          \and C.~Chiappini\inst{1} \and G.~R.~Davies\inst{2,3} \and Y.~P.~Elsworth\inst{2,3} \and B.~Mosser\inst{4}   \and M.~N.~Lund\inst{2,3} \and A.~Miglio\inst{2,3} \and W.~J.~Chaplin\inst{2,3}\and T.~S.~Rodrigues\inst{5,6} \and C.~Boeche\inst{7}   \and M.~Steinmetz\inst{1} \and G.~Matijevi\v c\inst{1} \and G.~Kordopatis\inst{1} \and J.~Bland-Hawthorn\inst{8} \and U.~Munari\inst{9} \and O.~Bienaym\'e\inst{10} \and K.~C.~Freeman\inst{11} \and B.~K.~Gibson\inst{12} \and G.~Gilmore \inst{13} \and E.~K.~Grebel\inst{7} \and A.~Helmi\inst{14} \and A. Kunder\inst{1} \and P.~McMillan\inst{15} \and J.~Navarro\inst{16} \and Q.~A.~Parker\inst{17} \and W.~Reid\inst{18,19} \and G.~Seabroke\inst{20} \and S.Sharma \inst{21} \and A.~Siviero\inst{6} \and F.~Watson\inst{22} \and R.~F.~G.~Wyse\inst{23} \and T.~Zwitter\inst{24} \and A.~Mott\inst{1}}

   \institute{Leibniz-Institut f\"ur Astrophysik Potsdam (AIP), An der Sternwarte 16, 14482 Potsdam, Germany
\and
School of Physics and Astronomy, University of Birmingham, Edgbaston, Birmingham, B15 2TT, UK
\and
Stellar Astrophysics Centre, Department of Physics and Astronomy, Aarhus University,  DK-8000 Aarhus C, Denmark 
\and
LESIA, Observatoire de Paris, PSL Research University, CNRS, Universit\'{e} Pierre et Marie Curie, Universit\'{e} Paris Diderot, 92195 Meudon, France
\and
Osservatorio Astronomico di Padova, INAF, Vicolo dell'Osservatorio 5, I-35122 Padova, Italy
\and
Dipartimento Fisica e Astronomia, Universit\'a di Padova, I-35122 Padova, Italy
\and
Astronomisches  Rechen-Institut,  Zentrum  f\"ur  Astronomie  der  Universit\"at  Heidelberg, M\"onchhofstr.  12-14,  D-69120  Heidelberg, Germany
\and
Sydney Institute for Astronomy, School of Physics, University of Sydney, NSW 2006, Australia
\and
Osservatorio Astronomico di Padova, INAF, I-36012 Asiago (VI), Italy
\and
Observatoire astronomique de Strasbourg, Université de Strasbourg, CNRS, UMR 7550, 11 rue de l’Université, F-67000 Strasbourg, France
\and
Research School for Astronomy and Astrophysics, Mount Stromlo Observatory, The Australian National University, ACT 2611, Australia
\and
E.A. Milne Centre for Astrophysics, University of Hull, Hull, HU6 7RX, United Kingdom
\and
Institute of Astronomy Cambridge University, Madingley Road Cambridge CB3 0HA, United Kingdom
\and
Kapteyn Astronomical Institute, University of Groningen, P.O. Box 800, 9700 AV Groningen, Netherlands
\and
Lund Observatory, Box 43, SE-221 00 Lund, Sweden
\and
Senior CIFAR Fellow, Dept. of Physics and Astronomy, University of Victoria, Victoria, BC, Canada V8P 5C2
\and
Department of Physics, The University of Hong Kong, Hong Kong, China
\and
Department of Physics and Astronomy, Macquarie University, Sydney, NSW 2109, Australia
\and
Centre for Astronomy, Astrophysics and Astrophotonics, Macquarie University, Sydney, NSW 2109, Australia
\and
Mullard Space Science Laboratory, University College London, Holmbury St Mary, Dorking, RH5 6NT, UK
\and
Sydney Institute for Astronomy, School of Physics, University of Sydney, NSW 2006, Australia
\and
Australian Astronomical Observatory, PO Box 915, North Ryde, NSW 1670, Australia
\and
Department of Physics and Astronomy, Johns Hopkins University, 3400 N. Charles St, Baltimore, MD 21218, USA
\and
Faculty of Mathematics and Physics, University of Ljubljana, 1000 Ljubljana, Slovenia
}

   \date{ accepted }

\abstract{We present a set of 87 RAVE stars with detected solar like oscillations, observed during Campaign 1 of the K2 mission (RAVE K2-C1 sample). This dataset provides a useful benchmark for testing the gravities provided in RAVE Data Release 4 (DR4), and is key for the calibration of the RAVE Data Release 5 (DR5).
The RAVE survey collected medium-resolution spectra (R=7,500) centred in the Ca II triplet (8600\AA) wavelength interval, which although being very useful for determining radial velocity and metallicity, even at low SNR, is known be affected by a \logg-\teff~degeneracy. This degeneracy is the cause of the large spread in the RAVE DR4 gravities for giants. 
The understanding of the trends and offsets that affects RAVE atmospheric parameters, and in particular \logg, is a crucial step in obtaining not only improved abundance measurements, but also improved distances and ages.
In the present work, we use two different pipelines, GAUFRE (Valentini et al. 2013) and Sp\_Ace (Boeche \& Grebel 2016), to determine atmospheric parameters and abundances by fixing  \logg~to the seismic one. Our strategy ensures highly consistent values among all stellar parameters, leading to more accurate chemical abundances. A comparison of the chemical abundances obtained here with and without the use of seismic \logg~information has shown that an underestimated (overestimated) gravity leads to an underestimated (overestimated) elemental abundance (e.g. [Mg/H] is underestimated by $\sim$0.25 dex when the gravity is underestimated by 0.5 dex). 
We then perform a comparison between the seismic gravities and the spectroscopic gravities presented in the RAVE DR4 catalogue, extracting a calibration for  \logg~of RAVE giants in the colour interval 0.50<($J - K_S$)<0.85. 
Finally, we show a comparison of the distances, temperatures, extinctions (and ages) derived here for  our RAVE K2-C1 sample with those derived in RAVE DR4 and DR5. DR5 performs better than DR4 thanks to the seismic calibration, although discrepancies can still be important for objects for which the difference between  DR4/DR5 and seismic gravities differ by more than $\sim$0.5~dex. The method illustrated in this work will be used for analysing RAVE targets present in the other K2 campaigns, in the framework of Galactic Archaeology investigations.}

   \keywords{surveys --
                stars: late-type --
                stars: oscillations --
		stars: fundamental parameters --
		stars: abundances --
		techniques: spectroscopic
               }

   \maketitle
%

\section{Introduction}

Galactic spectroscopic surveys play a key role in modern astrophysics. They provide large datasets of stellar atmospheric parameters, velocities, distances and abundances, making it possible to test modern models of Galactic dynamical and chemical evolution. RAVE \citep{Steinmetz2006}, the Gaia-ESO survey (\citealp{Gilmore2012,Randich2013}), GALAH \citep{deSilva2015}, APOGEE \citep{Majewski2015}, SEGUE \citep{Yanny2009}, and LEGUE \citep{Zhao2012} are providing stellar catalogues of several hundred thousand objects. 

Red giant stars are among the primary targets of spectroscopic Galactic surveys, since they are intrinsically bright and common objects and they can be observed in several components of the Milky Way. In addition, they cover a wide range in age, making it possible to reconstruct the history of our Galaxy. However, the measure of stellar atmospheric parameters (effective temperature, \Teff, and surface gravity, \logg) of red giants via spectroscopic analysis is affected by known systematics (\citealp{Morel2012, Heiter2015}). 

In this work we focus on the \logg~determination. It is a well-known problem in the literature that the \logg~for late type stars suffers from systematics of the order of 0.2 dex (\citealp{Morel2012, Heiter2015, Takeda2015, Takeda2016}). The causes of these systematics are numerous, and not only the different techniques adopted by different authors (e.g., ionization balance, line profile fitting). Among the culprits there are also the adoption of inaccurate line parameters (such as oscillator strenght), the assumption of Local Thermodynamical Equilibrium (LTE) and 1-D conditions, degeneracies, and  noisy or ill continuum-normalised spectra. As a consequence, an inaccurate measure of the gravity can lead to inaccurate estimates of \Teff, chemical abundances, distances and stellar age, since the determinations of these quantities are linked and ultimately dependent on the \logg~estimate. 

With the advent of asteroseismology and thanks to the valuable observations performed using the CoRoT \citep{Baglin2006} and \kepler \citep{Borucki2010} satellites, it has been possible to derive with high precision fundamental properties of Red Giant stars, such as mass (\M) and radius (\R) by using their global seismic properties \dnu (frequency separation) and \numax (frequency of maximum oscillation power). It was immediately realised that asteroseismology could have a large impact on galactic populations studies (Miglio et al. 2009, 2013). 

The surface gravity determined from stellar oscillations proved to be more precise and accurate than the one derived by using only spectroscopy (\citealp{Morel2012, Hekker2013, Heiter2015}). This seismic \logg, \logg$_{\rm seismo}$, can therefore be used as a powerful tool for testing the adopted spectroscopic pipelines and, eventually, calibrating them. In the recent years, pipelines that derive atmospheric parameters and abundances, implementing the seismic gravity, have been developed too, as GAUFRE \citep{Valentini2013}. Current spectroscopic surveys are largely taking advantage of the asteroseismic techniques, by including red giants for which asteroseismology is available, in their target list. CoRoT targets are now being observed by GES as calibrators \citep{Pancino2012}, \kepler targets have been used for calibrating APOGEE \citep{Pinsonneault2014} and LAMOST \citep{Wang2016} stellar surface gravities. The first results impacting Galactic Archeology using asteroseismology coupled with spectroscopy are now starting to appear (\citealp{Chiappini2015, Martig2015, Anders2016, Anders2016b}, Valentini et al. in prep.).  

The Kepler K2 mission (started on June 2014, Howell et al. 2014) is the continuation of the successful \kepler space mission. In 2014, the failure of two reaction wheels made the observations of the original field not feasible any more. For this reason a new mission, K2, was conceived, planning 80-day observational runs of a set of 14 fields located along the ecliptic plane. K2 is able to detect solar-like oscillations in field red giants \citep{Stello2015} and clusters \citep{Miglio2016}, and the light curves were of sufficient quality for measuring seismic parameters. The satellite is now observing several hundreds of RAVE targets, making it now possible to obtain asteroseismic informations also for RAVE red giants (\kepler, which field was in the north hemisphere, has no common target with RAVE, and the few RAVE targets in common with CoRoT have too noisy light curves).

The RAVE survey, completed in 2013, is the precursor of larger spectroscopic surveys. It provided an unprecedented view of our Galaxy, observing $\sim$ 500,000 targets in the southern hemisphere. The DR4 catalogue \citep{Kordopatis2013}, provides stellar velocities and atmospheric parameters plus metallicities, with special attention devoted to the derivation of reliable metallicities using calibration datasets. The database also contains seven element abundances (Mg, Al, Si, Ca, Ti, Fe and Ni), derived using a dedicated abundances pipeline \citep{Boeche2011}. The estimated errors in abundance, based on a comparison with reference stars, depend on the element and signal to noise ratio (hereafter SNR). For SNR>40 they range from 0.17 dex for Mg, Al and Si to 0.3 dex for Ti and Ni. The error for Fe is estimated as 0.23 dex. DR4 also provides distances, that were derived by using two different methods: via isochrone fitting \citep{Zwitter2010} and via Bayesian distance-finding with kinematic corrections \citep{Binney2013}. The later method also gives an estimate of the stellar ages, albeit with large uncertainties (see \citet{Binney2013} for a discussion). 
 
The \logg~determination is a problematic step for RAVE: its spectral interval suffers from a strong \logg-\teff~degeneracy, that causes an inaccurate \logg~measure for red giants and an offset, that causes the misplacement of the red clump of $\sim$ 0.3 dex \citep{Kordopatis2011,Kordopatis2013, Binney2013}. The main aim of this paper, as first in a series (were we will use K2 targets in common with RAVE for galactic archaeology purposes) is to show the impact of using the precise and accurate seismic gravity in the outcome temperatures and abundances of RAVE targets. We also show how the approach discussed here helps improving the RAVE stellar parameters and abundances. As shown in \citet{Bruntt2012}, \citet{Thygesen2012}, and \cite{Morel2014} asteroseismology can play an important role in this respect, as it provides precise and accurate gravities, once more helping to break remaining degeneracies. Additional improvements regarding the lifting of the degeneracy are shown in DR5 \cite{Kunder2016}, by using the new APASS photometric information, the Infra-Red Flux Method, and the \logg~calibration presented in this work. 

The paper is organised as follows: in Section 2, we present the RAVE targets that have been observed in K2 Campaign 1; in Section 3 we present the seismic data available for our sample; in Section 4 we describe our spectroscopic analysis strategy in order to obtain highly consistent stellar parameters and therefore accurate stellar abundances for our sample. In Section 5 we compared our results with those of RAVE DR4 for the same stars, providing a calibration for the \logg$_{\rm RAVE~DR4}$. Section 6 focus in showing how variations in \logg~impacts elements abundances, and what is the safe parameter space over which our calibration can be applied. Distances, reddening (and ages), determined via a Bayesian approach using asteroseismology and the newly determined atmospheric parameters are shown in Section 7. In this section we also provide a comparison with the values obtained in DR4 an DR5 for the same stars. In particular, DR5 has made use of the seismic analysis presented in this work. In Section 8, we summarise our results.  

\begin{figure}
   \includegraphics[width=0.99\columnwidth]{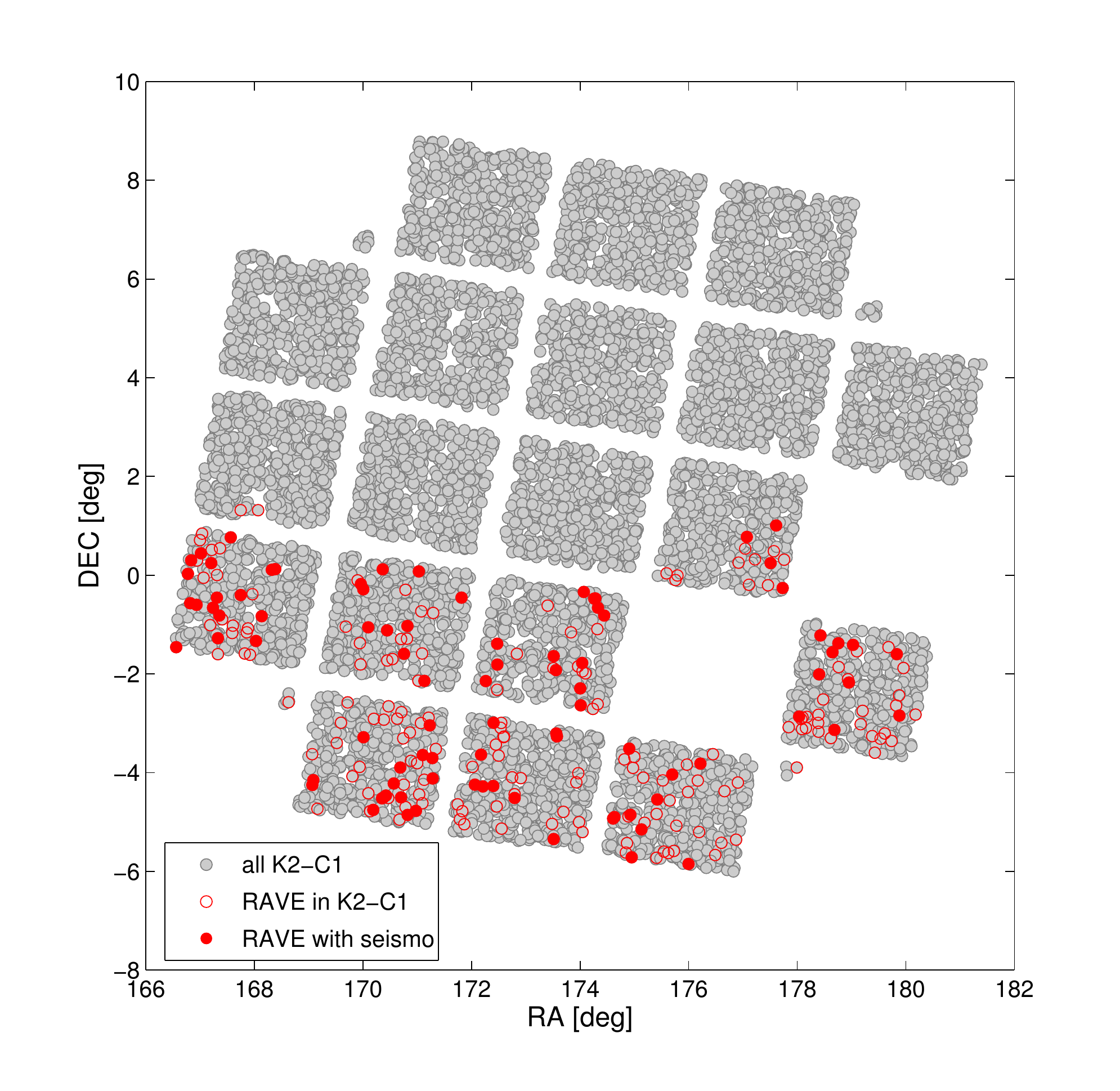}
   \caption{RA-DEC position of the targets observed by K2 during Campaign 1 (grey dots), the field is centred at 11:35:46 	+01:25:02 and it was observed from 30-05-2014 to	21-08-2014. Empty red circles mark the RAVE stars observed by K2, while full red circles mark the 87 RAVE targets with detected oscillations.}
              \label{Fig:field}%
\end{figure}

\begin{figure}
   \includegraphics[width=0.99\columnwidth]{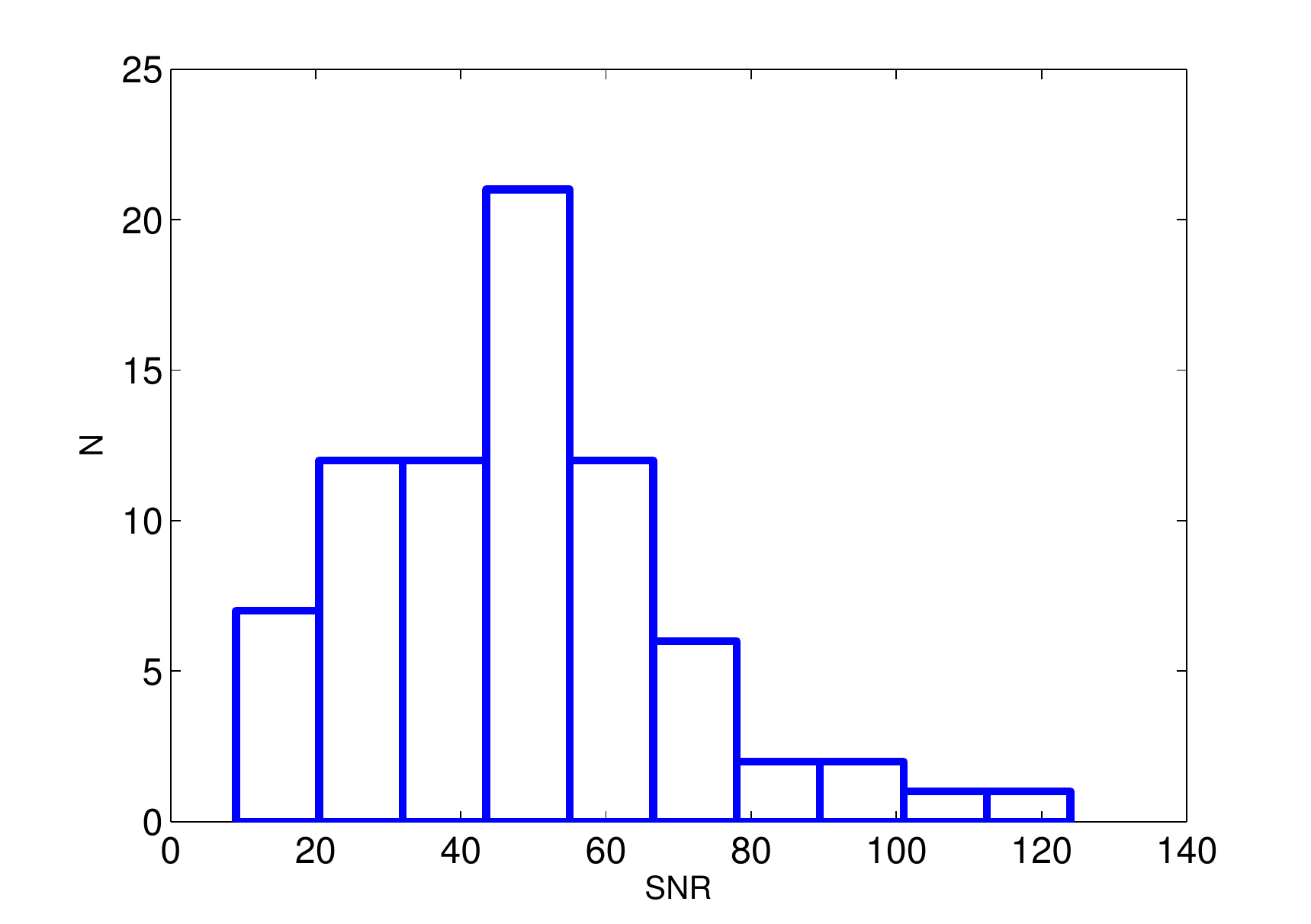}
   \caption{SNR distribution of the spectra of the 87 RAVE stars possessing asteroseismology.}
              \label{Fig:SNR}%
\end{figure}


\section{RAVE targets in K2 Campaign 1}

The K2 Campaign 1 is a field 100  deg$^2$ large, centred at RA 11~h 35~m 46~s DEC +01$^\circ$ 25' 02'' (l$=$265, b$=$$+$58), it is thus a field almost perpendicular with respect to the Galactic Plane.

\begin{figure}
   \includegraphics[width=0.99\columnwidth]{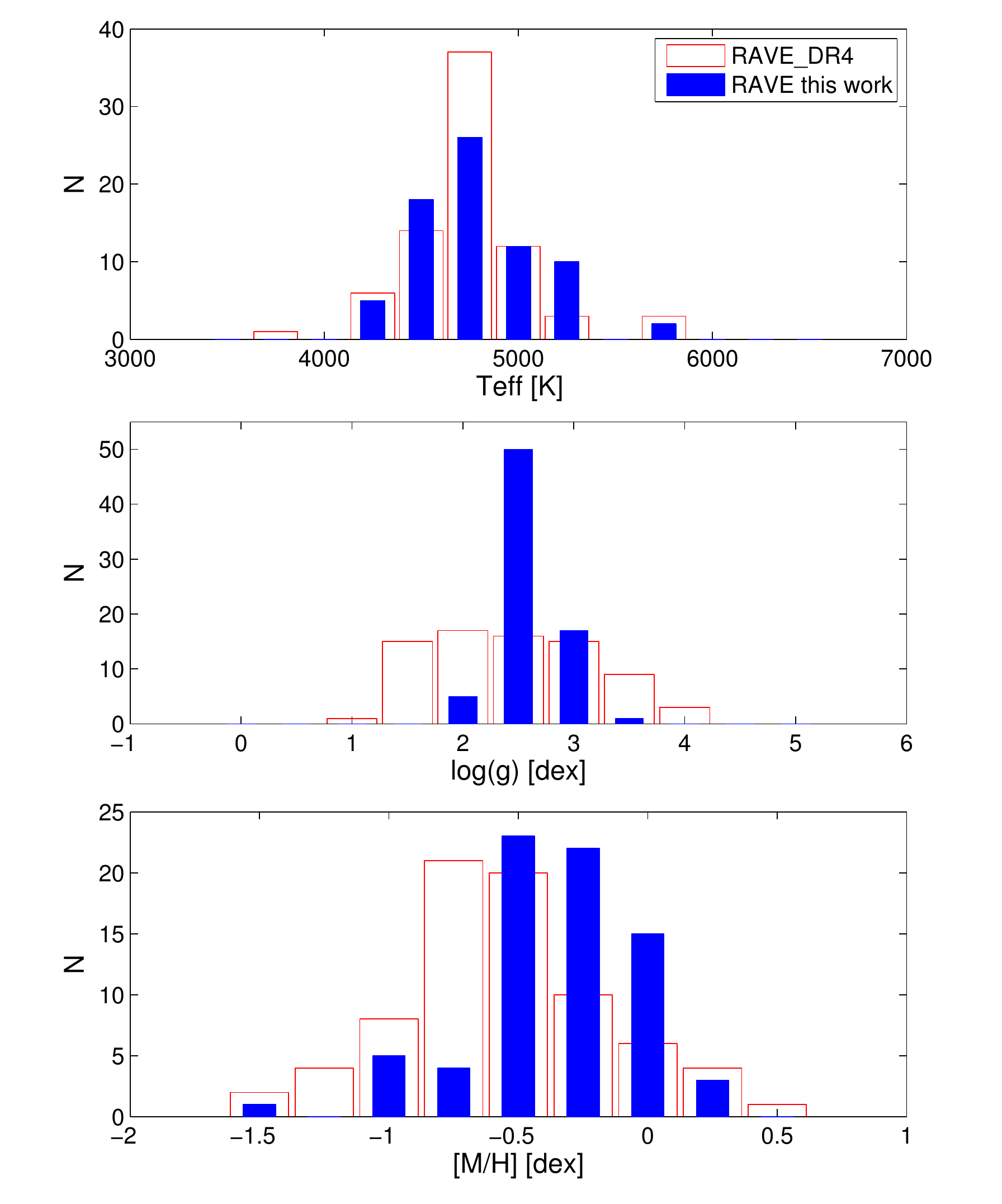}
   \caption{Distribution of \teff,\logg~and [M/H] of the RAVE stars analysed in this work. RAVE DR4 values are plotted in empty, red bars; the new atmospheric parameters derived by using asteroseismology are plotted with blue, filled bars.}
              \label{Fig:hist}%
\end{figure}

\begin{figure*}
   \includegraphics[width=1\textwidth]{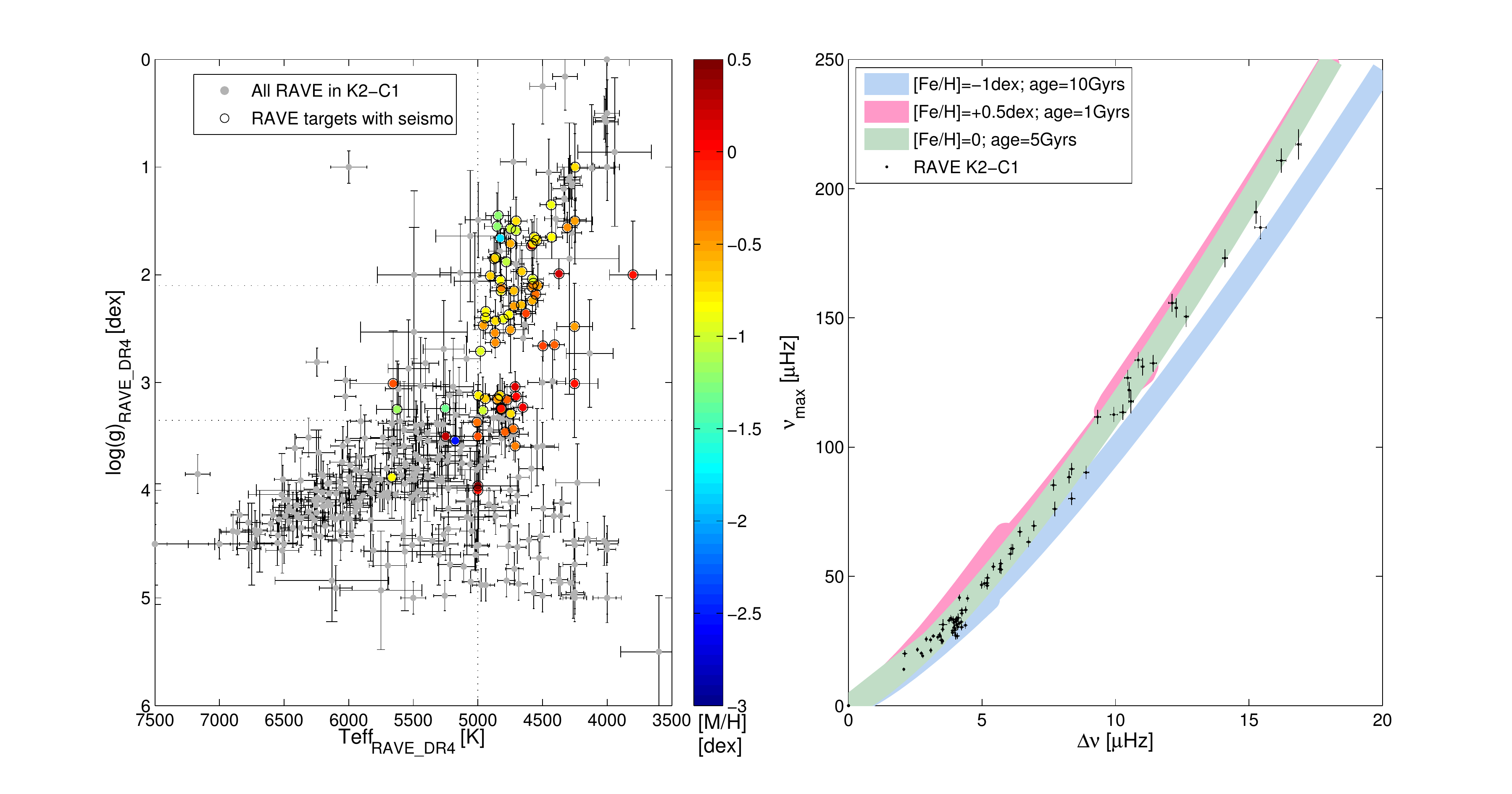}
   \caption{Left panel: the \logg-\teff~distribution of RAVE targets in K2-C1 target list (grey dots). Atmospheric parameters and errors are taken from RAVE DR4. Targets possessing \dnu and \numax are colour enhanced (by following calibrated [M/H] from RAVE DR4 catalogue) and circled in black. The dashed lines in \logg~mark the K2 detection limits at 2.1 and 3.35 dex. Right panel: \dnu and \numax distribution the 87 RAVE targets in K2-C1 possessing seismic parameters. The distribution is superimposed on three \dnu-\numax distributions calculated following Padova isochrones, taken at three different metallicities and ages.}
              \label{Fig:distrKa}%
\end{figure*}

In the field of view of K2 Campaign 1 there are 1400 RAVE targets, among those 247 are present in the K2-C1 target list (see Fig.~\ref{Fig:field}). Seismic parameters \deltanu and \numax have been measured for 87 objects (see Sec.~3 for details). The SNR, \teff, \logg~and [M/H] distributions are shown in Fig.~\ref{Fig:SNR} and Fig.~\ref{Fig:hist} (red histogram), while the \logg-\teff diagram of the targets, constructed using DR4 data, is shown in the left panel of Fig.~\ref{Fig:distrKa}. As visible in the last panel of Fig.~\ref{Fig:hist}, the metallicity distribution, computed using asteroseismology (filled blue histogram), tends to be more metal-rich than the RAVE DR4 one (red empty histogram), but it covers a large metallicity interval.   

\subsection{Spectra}

RAVE spectra were taken using the 6dF facility, a multi-fiber spectrograph  mounted at the 1.2-m UK Schmidt Telescope of the Australian Astronomical Observatory (AAO). Spectra cover a wavelength range of $\sim$400 \AA, from 8410 \AA ~to 8795 \AA. RAVE resolution is $R=\Delta\lambda/\lambda=7,500$. This wavelength range is widely used in the field of Galactic Archaeology: the presence of the strong Ca II triplet ($\lambda$ = 8498.02 \AA, 8542.09 \AA, 8662.14 \AA) makes it possible to measure radial velocity (RV) even at low SNR. The Ca II triplet acts as metallicity indicator too(e.g. \citet{DaCosta1998}). Using several features of Fe and $\alpha$-elements (Mg, Si, Ca, Ti), it is also possible to measure element abundances, as shown in \citet{Boeche2011}, \citet{Kordopatis2013}, \citet{Boeche2014}. The same wavelength interval is covered by Gaia-ESO survey (HR21 set-up of the FLAMES-GIRAFFE multi-object spectrograph) and the Gaia Radial Velocity Spectrometer (Gaia-RVS).

\subsection{Photometry}

RAVE DR4 catalogue contains DENIS DR3 \citep{DENIS} and 2MASS \citep{Cutri2003} photometry. In this work the photometry of DR4 is implemented with the APASS photometry for RAVE targets from \citet{Munari2014}. APASS provided photometry in the Landolt {\it BV} and Sloan {\it g'r'i'} bands. APASS photometry is available for all the 87 targets of our RAVE-K2 sample in Campaign 1. We have also added the $WISE$~$W$1 and $W$2 filters photometry, from the AllWISE Catalog \citep{Cutri2013}.

The  \citet{Munari2014} catalogue provides also photometric temperatures, computed in 6 different ways. For our analysis we focused on the \teff~derived by simultaneously fitting \EBV, in order to avoid systematics introduced by the adoption of a fixed value for distance, reddening, \logg~or [M/H].

\section{Asteroseismic data}
\label{Sec:seismo}

The Campaign 1 field was observed by K2 from May 30 2014 to August 21 2014. The satellite observed 21,647 targets in the field. 

RAVE targets analysed in this work were observed as part of the ``The K2 Galactic Archaeology Program Campaign 1''  (C1 proposal GO1059, \cite{Stello2015}). The target list of this project was composed of red giants belonging to some of the most important spectroscopic surveys, as RAVE, APOGEE and GALAH.

Pixel masks for the individual C1 targets were defined using the K2P$^2$ pipeline (K2-Pixel-Photometry; \citet{Lund2015}). First a summed image (over time) is constructed that includes the apparent motion of the stars on the CCD due to the characteristic $~6$ hour drift of the spacecraft (\citealp{Howell2014,VanCleve2016}). A set of unsupervised machine learning techniques are applied in K2P$^2$  to the summed image to define the pixel masks from which raw light curves are extracted. Instrumental features in the raw flux light curves are corrected for using the strong correlation of these with the stellar position on the CCD. Finally, the light curves are corrected for further artefacts using the KASOC filter \citep{HandbergLund2014} with adopted time scales of $\tau_{\rm long} = 3$ days and $\tau_{\rm short} = 0.25$ days for the median filters (we refer to \citet{HandbergLund2014} for additional information on the KASOC filter).

To estimate the frequency of maximum oscillation power we adopted the technique described in \citet{Davies2016}, based on fitting a background model to the data. We fitted model H of \citet{Kallinger2014}, comprised of two Harvey profiles, a Gaussian oscillation envelope, and an instrumental noise background. For the estimate of \numax we took the central frequency of the Gaussian component. We used  the median and the standard deviation to summarize the normal-like posterior probability density for \numax. The latter parameter has been measured for 87 RAVE stars. As an external check, \numax has been also estimated using the technique of \cite{Mosser2011}: the \numax values measured by the two independent techniques agree very well, with a median fractional difference below 1\%.

To estimate the average frequency separation, we adopted the method described in \citet{Mosser2009} and \citet{Mosser2011}. This method uses the expected frequency pattern of a Red Giant for identifying oscillations modes. \dnu was then measured for 86 RAVE stars. We performed a reliability check of the seismic parameters by using the PARSEC set of isochrones \citep{Bressan2012}, following the approach adopted in Valentini et al. (GES-in prep.). We considered 3 isochrones,at \FeH$=-$1.0, 0.0, $+$0.5 dex and of age 10, 5 and 1 Gyr respectively. All the 86 stars possessing both \dnu and \numax fall within the predicted distribution. 

In this work we therefore adopted the \numax and its uncertainty for 87 stars, measured using the \citet{Davies2016} technique. Of those stars with detected \numax, 86 possess \dnu values measured using \citep{Mosser2011} method (with their uncertainties).

\section{Spectroscopic analysis}
\label{Sec:SpAnalys}

\begin{table*}
\caption{Input physics of GAUFRE and Sp\_Ace codes.}            
\label{Tab:codes}      
\centering          
\begin{tabular}{lcccc}     
\hline\hline       
Code & Model Atmospheres  & Line parameters & Line formation code & Microturbulence\\ \hline
          &        &    &          &          \\                
GAUFRE        & \citet{Castelli2004} & VALD3 &  Synth3$^{(1)}$ & fixed (2 Km/s) \\
Sp\_Ace & \citet{Castelli2004} & VALD3 (refined)$^{(2)}$ & GCOG &  function of \Teff~and \logg$^{(3)}$ \\
          &        &    &          &          \\ 
\hline  
          &        &    &          &          \\              
\end{tabular}
\tablebib{
      (1) See \citealp{Kochukhov2007,Kochukhov2010}; (2) For details refer to Section~4 of \citet{Boeche2016}; (3) For details refer to Appendix~1 of \citet{Boeche2016}.
    }
\end{table*}

As widely discussed in Kordopatis et al. (2011, 2013), the wavelength interval observed by RAVE suffers from a strong degeneracy between effective temperature and surface gravity, because of the low resolution (R$\leq$10,000) combined with the small wavelength coverage. The wavelength interval possesses too few spectral features sensitive to \teff~or \logg~only, often the same feature is used as an \teff~and \logg~indicator at the same time. This leads to degeneracies, due to the fact that a spectral line can have the same depth and shape for two stars with different atmospheric parameters. A solution might be to identify additional spectral features sensitive to one parameter only, to change the algorithm in the pipeline (overcoming the classical $\chi^2$ minimization technique) or to use external information that already provides an indication about the temperature and gravity of the object.

In the work of RAVE DR4, the \logg-\teff ~degeneracy was partially solved by adopting a combination of a decision-tree algorithm and a projection method (method explained in \citet{Kordopatis2011}), with a rough initial \teff~selection based on photometric temperatures. 

In this work we show that, when seismic information is available, the determination of reliable atmospheric parameters and abundances is possible also for algorithms that use the distance minimization. The main problem with pipelines that adopt the minimum distance method, is that the degeneracies wipe out the identification of the minimum. For example, the pipeline risks converging at a secondary minimum, or that two very close secondary minimums merge, leading to a wrong and imprecise solution. Asteroseismology, combined with photometry and spectroscopy, avoids this problem: the \logg~is fixed to the seismic value and the temperature provided by photometry is used as prior, removing the degeneracy of the spectroscopic analysis, and reducing the risk of convergence into secondary minima.

For the spectroscopic analysis of the RAVE spectra we used two pipelines: GAUFRE, for the \logg~and \teff~determination, and Sp\_Ace for the determination of overall metallicity and abundances. Sp\_Ace had been already successfully used in previous tests for deriving stellar parameters and element abundances and performs well at low resolutions. GAUFRE works using seismic values, in order to iteratively derive \logg, \teff and \FeH. We decided for the adoption of two pipelines because, at the moment, Sp\_Ace does not allow the adoption of probabilistic priors, but it takes fixed \teff~and \logg~as input. The two pipelines are described in the following  Subsect.~\ref{Sec:pipelines}.

\subsection{Description of the adopted spectroscopic pipelines}
\label{Sec:pipelines}

{\bf GAUFRE}

GAUFRE \citep{Valentini2013} is a spectroscopic pipeline that implements asteroseismology in the derivation of atmospheric parameters. GAUFRE pipeline is currently used in the analysis of CoRoT-GES targets, Valentini et al. (GES-in prep.). 
 
It is a C++ collection of several routines, designed for the spectroscopic analysis of high-resolution spectra of F-G-K giants in the optical domain. For the analysis of the RAVE spectra we used the GAUFRE-SISMO and the GAUFRE-CHI2 routines, to iteratively derive atmospheric parameters via $\chi^2$ fitting on a library of synthetic spectra, by fixing the gravity to the seismic one. The spectral library used in this work, is the one provided by L. Fossati and degraded to the RAVE resolution of R=7,500 and covering the 8350-8850 \AA~ spectral range. The synthetic  spectra has been computed using Synth3 code (\citealp{Kochukhov2007,Kochukhov2010}), using \citet{Castelli2004} model atmospheres and VALD3 linelist. Synthetic spectra were renormalized using the same function as most of the RAVE spectra: an order 4 cubic spline with 1.5$\sigma$ and 3.0$\sigma$ low- and high-level rejection thresholds (\citealp{Zwitter2008,Siebert2011}). For our analysis we masked the cores of the strong CaII triplet lines (that may be affected by NLTE effects, see \citet{Jorgensen1992}).

In this work the GAUFRE pipeline has been used for iteratively deriving \teff~and \logg, by using APASS photometric temperatures as a prior (providing a flexibility of $\pm$500 K) and fixing the gravity to the seismic one. The validation of this method is discussed in Subsec.~\ref{Sec:Validation}. The input physics of the two codes is summarized in Table~\ref{Tab:codes}.

{\bf Sp\_Ace}

\Space is a FORTRAN95 code that can estimate \teff, \logg, and
elemental abundances from normalized, radial velocity corrected stellar
spectra. It derives the parameters seeking the minimum $\chi^2$ computed
from the observed spectrum and model spectra, the latter constructed by
\Space from a library of General Curves-Of-Growth (GCOG, see \citet{Boeche2016}
for details). As shown in \citet{Boeche2016} \Space~~  
performs well between spectra resolutions 2\,000 and 20\,000, which include the RAVE
resolution.  Among other features, this code allows the user to determine
the chemical abundances by fixing \logg~and/or
\teff~to trusted values. In this work we run \Space by adopting the options ``ABD\_loop'' (which rules the \Space internal
iterations between the routines that estimates the stellar parameters \teff~and \logg~and the abundances) and ``norm\_rad = 10'' (which rules the re-normalization of the observed spectrum).

We used the Sp\_Ace pipeline for determining metallicity and abundances, by fixing the \teff~and \logg~to the values derived iteratively by GAUFRE. The validation of this method is discussed in Subsec.~\ref{Sec:Validation}. 

\subsection{Pipelines validation}
\label{Sec:Validation}
\begin{figure*}
   \includegraphics[width=1\textwidth]{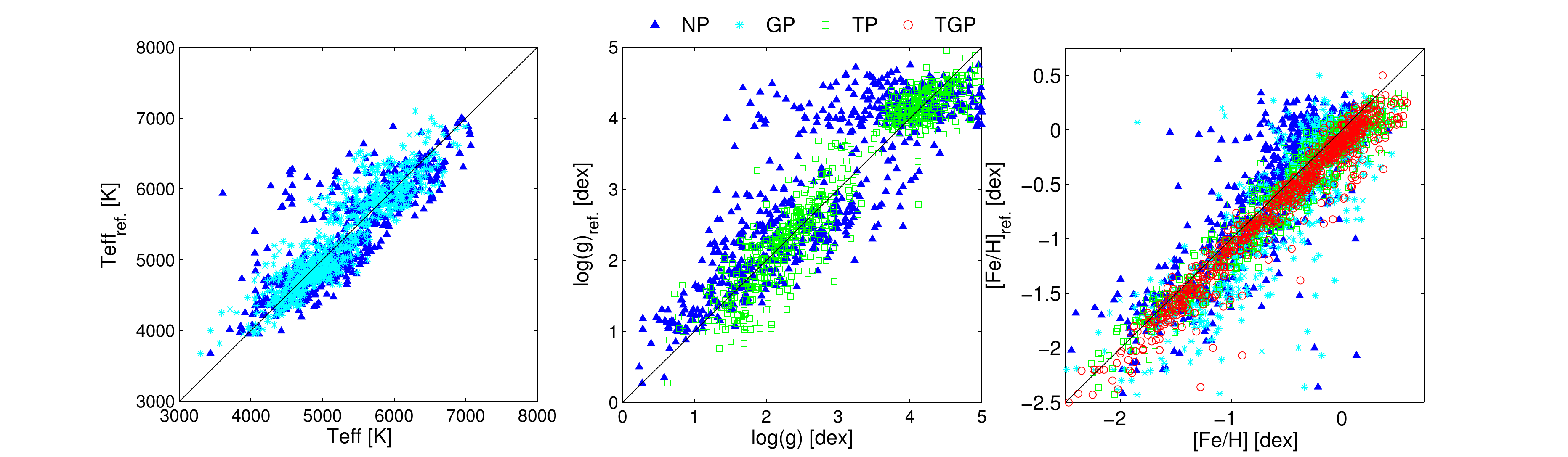}
   \caption{Comparison of the atmospheric parameters (from left to right: \teff, \logg~and \FeH) of the calibration set versus the values derived by using the GAUFRE  pipeline. Parameters were derived adopting four different strategies, following the same code as Fig.~\ref{Fig:TGM_space}. Mean dispersions and offsets are displayed in Table~\ref{Tab:TGM_gaufre}.}
              \label{Fig:TGM_gaufre}%
\end{figure*}
\begin{figure*}
   \includegraphics[width=1\textwidth]{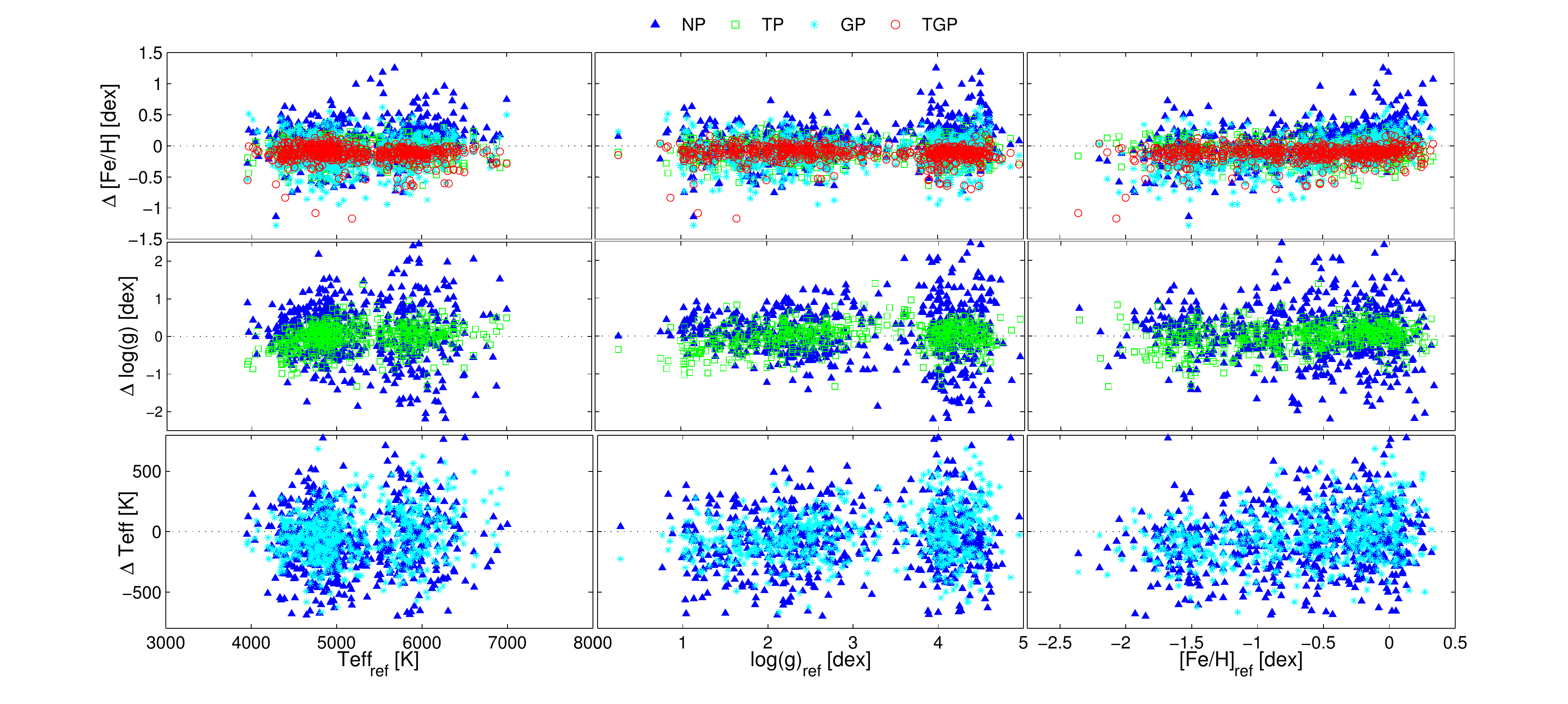}
   \caption{Comparison of the atmospheric parameters (from top to bottom: \FeH, \logg~and \Teff) measured by the GAUFRE pipeline, versus the atmospheric parameters available in literature. Different symbols mark different approaches, following the same code as Fig.~\ref{Fig:TGM_space}. Mean dispersions and offsets are displayed in Table~\ref{Tab:TGM_gaufre}.}
              \label{Fig:Trends_gaufre}%
\end{figure*}

For the validation of the two pipelines, we derived atmospheric parameters and abundances for the set of reference stars used in RAVE DR4 \citep{Kordopatis2013}. The RAVE DR4 calibration datasets of observed spectra consist in 809 spectra of giants and dwarfs belonging to the field or Open Clusters. All the spectra have a SNR $>$ 40 pixel$^{-1}$. The sample was constructed in order to cover as much as possible the parameter space of the stars observed by the RAVE survey. RAVE reference catalogue comprises heterogeneous sources: a set of 169 RAVE giants and dwarfs with multiple PASTEL entries \citep{Soubiran2010}, 224 dwarfs and giants present in the CFLIB library \citep{Valdes2004}, 163 giants observed by Fulbright et al. (in prep.), 229 spectra of giants and dwarfs from \citet{Ruchti2011}, 22 spectra of stars belonging to M67 and IC 4651 open clusters \citep{Pancino2010,Pasquini2004}, and two spectra of the metal poor (\FeH=$-$4.2) giant CD-38245 \citep{Cayrel2004}. For details regarding the construction and the computation of the atmospheric parameters of the RAVE calibration data sets, refer to \citet{Kordopatis2013}. 

In order to simulate what happens using asteroseismology and when fixing different parameters, we run the two pipelines on the calibration set in 4 different ways:
\begin{itemize}
\item no constraints in \logg~nor \teff~(coded as -NP);
\item \teff~fixed (coded as -TP) ;
\item \logg~fixed (coded as -GP);
\item fixed \teff~and \logg~(coded as -TGP). 
\end{itemize}

\begin{figure*}
   \includegraphics[width=1\textwidth]{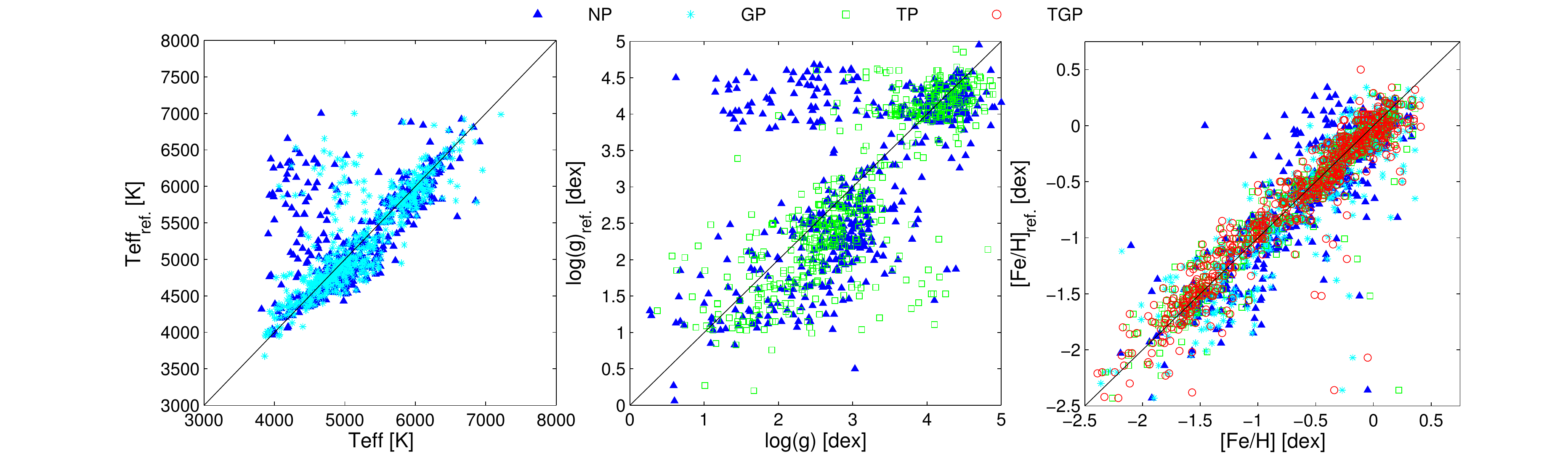}
   \caption{Comparison of the atmospheric parameters (from left to right: \teff, \logg~and \FeH) of the calibration set versus the values derived by using \Space pipeline. Parameters were derived adopting four different strategies: with no prior (blue triangles), fixing the temperature to the real value (green squares), fixing the gravity (cyan stars) and fixing temperature and gravity (red circles). Mean dispersions and offsets are displayed in Table~\ref{Tab:TGM_space}.}
              \label{Fig:TGM_space}%
\end{figure*}

\begin{figure*}
   \includegraphics[width=1\textwidth]{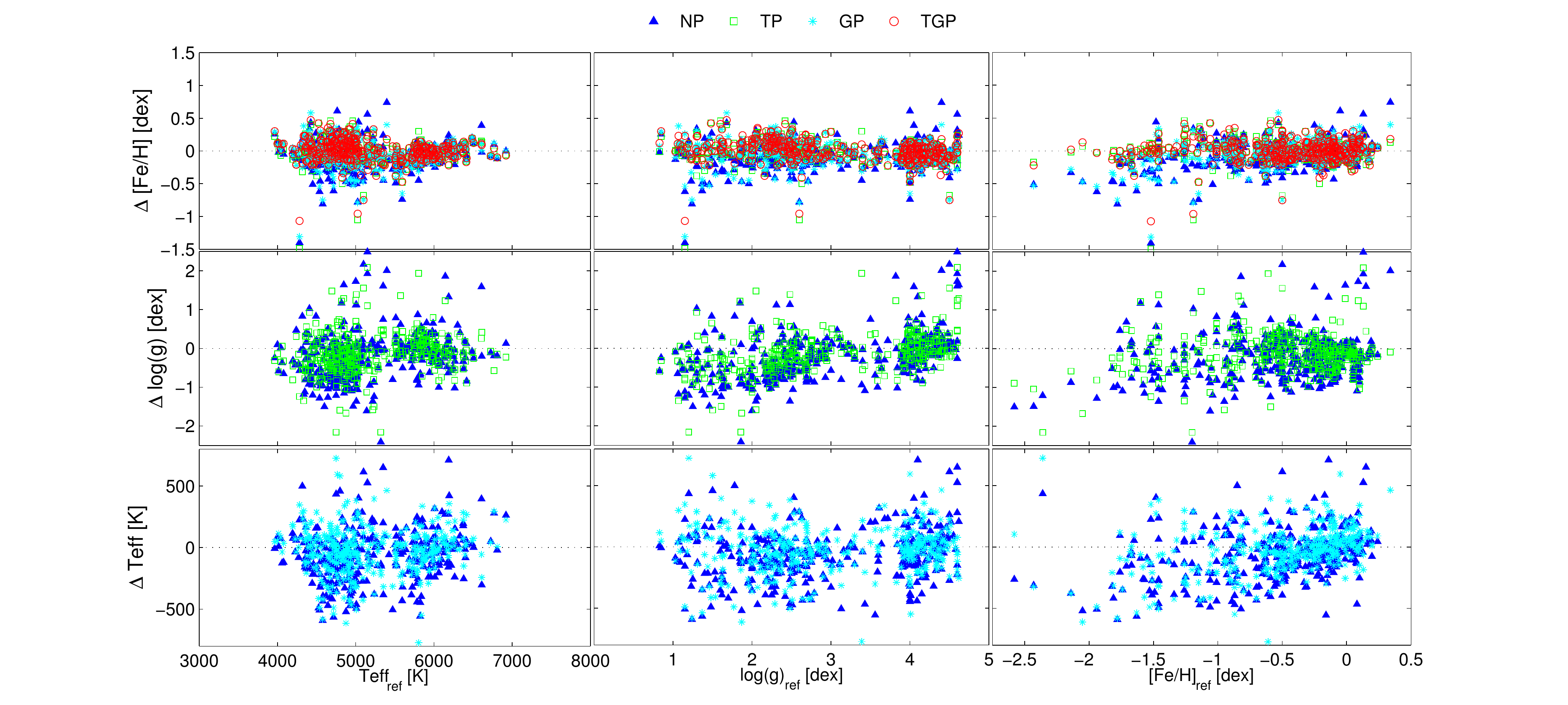}
   \caption{Comparison of the atmospheric parameters (from top to bottom: \FeH, \logg~and \Teff) measured by the \Space pipeline, versus the atmospheric parameters available in literature. Different symbols mark different strategies adopted for the analysis, following the same code as in Fig.~\ref{Fig:TGM_space}. Mean dispersions and offsets are displayed in Table~\ref{Tab:TGM_space}.}
              \label{Fig:Trends_space}%
\end{figure*}

\begin{figure*}
   \includegraphics[width=1\textwidth]{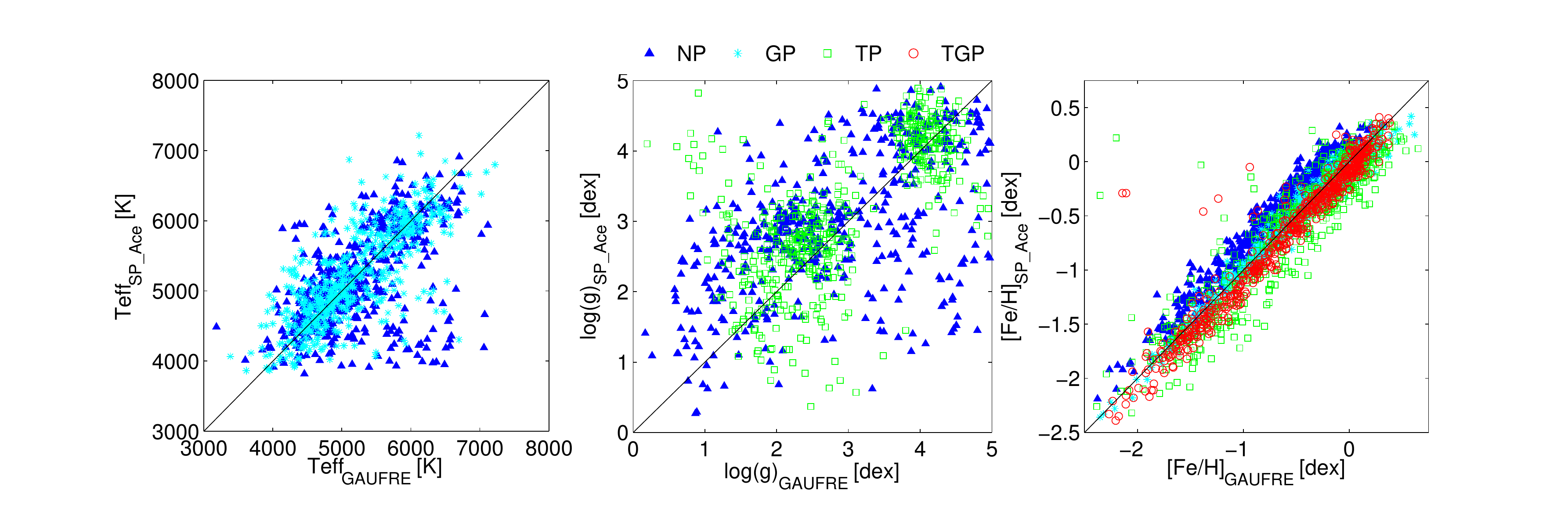}
   \caption{Comparison of the atmospheric parameters (from left to right: \teff, \logg~and \FeH) of the calibration set atmospheric parameters as measured by GAUFRE versus the values derived with the \Space pipeline. Parameters were derived adopting four different strategies: with no prior (blue triangles), fixing the temperature to the real value (green squares), fixing the gravity (cyan stars) and fixing temperature and gravity (red circles). Mean dispersions and offsets are displayed in Table~\ref{Tab:TGM_space_gaufre}.}
              \label{Fig:TGM_space_gaufre}%
\end{figure*}

\begin{figure*}
   \includegraphics[width=1\textwidth]{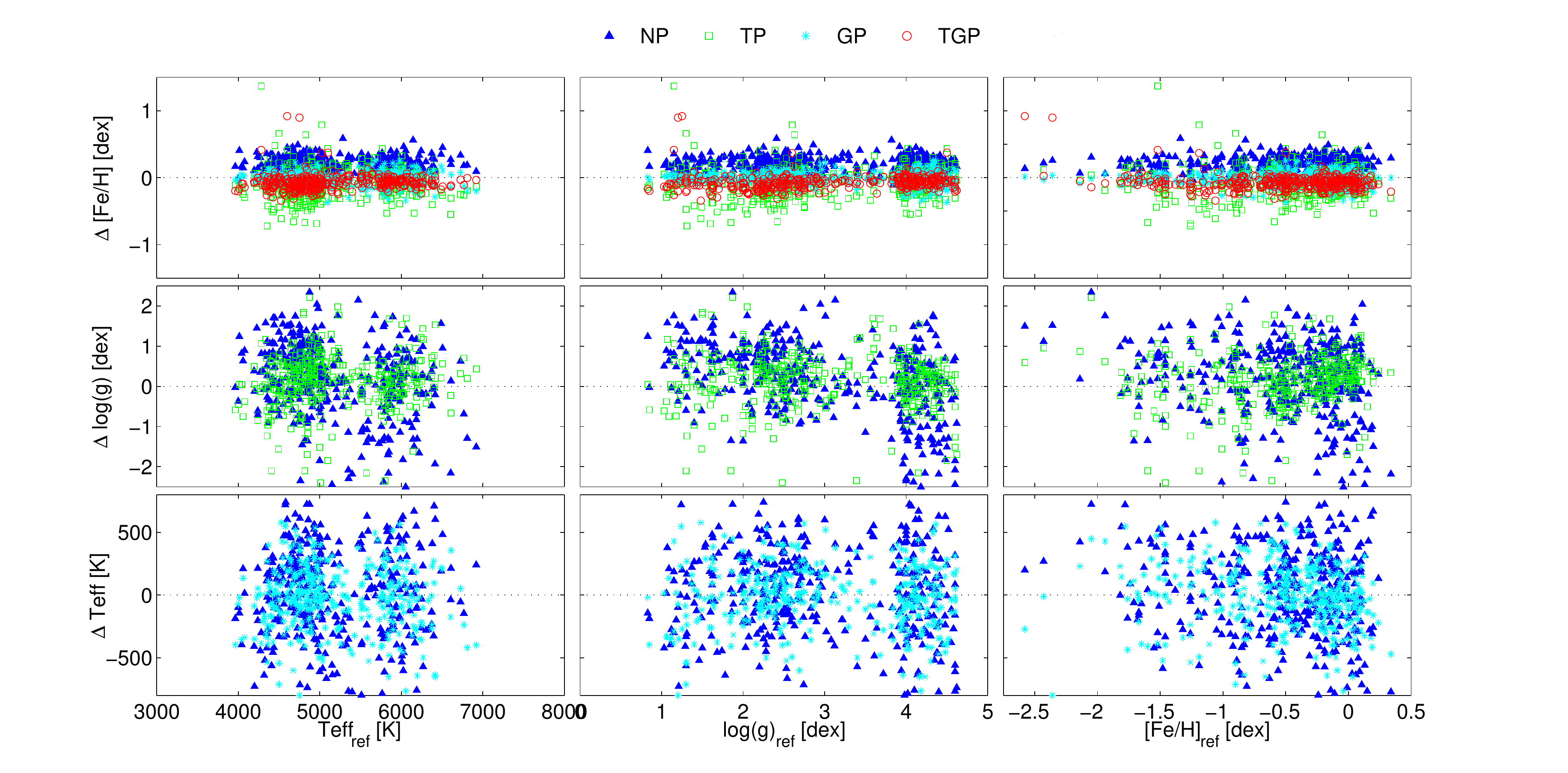}
   \caption{Comparison of the atmospheric parameters (from top to bottom: \FeH, \logg~and \Teff) measured by the \Space pipeline, versus the atmospheric parameters measured by GAUFRE. Different symbols mark different strategies adopted for the analysis, following the same code as in Fig.~\ref{Fig:TGM_space_gaufre}. Mean dispersions and offsets are displayed in Table~\ref{Tab:TGM_space_gaufre}.}
              \label{Fig:Trends_space_gaufre}%
\end{figure*}

Due to the limits of both pipelines, we considered only those targets with  \FeH$>$ $-$2.5 dex. The comparisons of the reference literature values with those derived by the two pipelines are shown in Fig. \ref{Fig:TGM_space} and Fig. \ref{Fig:TGM_gaufre}  for \teff, \logg~and \FeH. Offsets and dispersions of each pipeline, for all the 4 runs, are shown in Table \ref{Tab:TGM_space} and Table \ref{Tab:TGM_gaufre}. Possible trends and offsets have been investigated in Fig. \ref{Fig:Trends_space} and Fig. \ref{Fig:Trends_gaufre}, for the \Space and GAUFRE pipelines respectively. In the red giant regime, in the -NP analysis, the two pipelines show an offset in \logg~and a large spread, plus a trend that persists also when fixing the temperature to the literature value. 
A direct comparison between the \Space and GAUFRE pipeline is illustrated in Fig.~\ref{Fig:TGM_space_gaufre} and Fig.~\ref{Fig:Trends_space_gaufre}, while offsets and dispersions are reportedin Table~\ref{Tab:TGM_space_gaufre}. These offsets, dispersions and trends are the result of the short wavelength coverage of the survey: in the 400 \AA~ of the spectrum there are insufficient identified features able to solve the \logg-\teff~degeneracy.

When the information on \logg~and \teff~are available, however, the two pipelines are capable of determining a value for metallicity that is in good agreement with the literature. The GAUFRE pipeline shows an offset in \FeH of $\sim$$-$0.10 dex, due to the presence of a strong feature in the synthetic spectra that are not present (or that are less strong) in the real spectra. This metallicity shift does not depend on any other atmospheric parameter and it can be corrected by adding $+$0.1 dex to the \feh value given by the pipeline, or by upgrading the linelist, correcting the line parameters of the problematic features. For this work we used only the \logg~and \teff~determined by GAUFRE, and we computed the overall metallicity and abundances using \Space. When fixing the \logg~and \teff~to the literature values, \Space shows no offset in metallicity ($-$0.01 dex for giant stars). Such hybrid use of results does not introduce internal inconsistencies. 

\begin{table}
\caption{Mean dispersions and offsets for \teff, \logg~and \feh~ of the GAUFRE pipeline with respect to the literature values for the calibration data set.}             
\label{Tab:TGM_gaufre}      
\centering          
\begin{tabular}{lc c| c c| c c}     
\hline\hline       
 & \multicolumn{2}{l|}{Teff [K]}  & \multicolumn{2}{l|}{log(g) [dex]} & \multicolumn{2}{l}{\FeH [dex]}\\
       & offset & $\sigma$ & offset & $\sigma$ & offset & $\sigma$ \\ \hline
          &        &    &          &      &         &      \\                 
NP        & $-123$ & 481 &  0.07 & 0.73 & 0.03 & 0.39 \\
NP-giants & $-159$ & 349 &  0.07 & 0.51 & $-0.05$ & 0.42 \\
TP        & -- & -- & 0.21 & 0.33 & $-0.10$ & 0.19 \\    
TP-giants & -- & -- & 0.13 & 0.35 & $-0.10$ & 0.17 \\         
GP        & $-138$ & 241 & -- & -- & $-0.15$ & 0.35 \\
GP-giants & $-123$ & 287 & -- & -- & $-0.07$ & 0.41 \\
TGP       & -- & -- & -- & -- & $-0.12$ & 0.20 \\
TGP-giants& -- & -- & -- & -- & $-0.11$ & 0.20 \\
\hline               
\end{tabular}
\end{table}

\begin{table}
\caption{Mean dispersions and offsets for \teff, \logg~and \feh~ of the SPACE pipeline with respect to the literature values for the calibration data set.}            
\label{Tab:TGM_space}      
\centering          
\begin{tabular}{lc c| c c| c c}     
\hline\hline       
 & \multicolumn{2}{l|}{Teff [K]}  & \multicolumn{2}{l|}{log(g) [dex]} & \multicolumn{2}{l}{\FeH [dex]}\\
       & offset & $\sigma$ & offset & $\sigma$ & offset & $\sigma$ \\ \hline
          &        &    &          &      &         &      \\                
NP        & $-52$ & 216 &  $-0.20$ & 0.64 & $-0.11$ & 0.39 \\
NP-giants & $-102$ & 186 &  $-0.40$ & 0.54 & $-0.15$ & 0.49 \\
TP        & -- & -- & $-0.17$ & 0.52 & $-0.05$ & 0.39 \\    
TP-giants & -- & -- & $-0.26$ & 0.52 & $-0.04$ & 0.49 \\         
GP        & $-25$ & 191 & -- & -- & $-0.07$ & 0.36 \\
GP-giants & $-46$ & 207 & -- & -- & $-0.07$ & 0.45 \\
TGP       & -- & -- & -- & -- & $-0.03$ & 0.36 \\
TGP-giants& -- & -- & -- & -- & $-0.01$ & 0.44 \\
\hline               
\end{tabular}
\end{table}


\begin{table}
\caption{Mean dispersions and offsets for \teff, \logg~and \feh~ of the GAUFRE pipeline with respect to the values measured by \Space for the calibration data set.}            
\label{Tab:TGM_space_gaufre}      
\centering          
\begin{tabular}{lc c| c c| c c}     
\hline\hline       
 & \multicolumn{2}{l|}{Teff [K]}  & \multicolumn{2}{l|}{log(g) [dex]} & \multicolumn{2}{l}{\FeH [dex]}\\
       & offset & $\sigma$ & offset & $\sigma$ & offset & $\sigma$ \\ \hline
          &        &    &          &      &         &      \\                
NP        & 25 & 380 &  $-0.01$ & 0.69 & 0.09 & 0.40 \\
NP-giants & $51$ & 242 &  0.09 & 0.54 & 0.05 & 0.48 \\
TP        & -- & -- & 0.00 & 0.31 & $-0.10$ & 0.16 \\    
TP-giants & -- & -- & $-0.07$ & 0.32 & $-0.17$ & 0.16 \\         
GP        & $-42$ & 176 & -- & -- & $-0.07$ & 0.38 \\
GP-giants & $-44$ & 153 & -- & -- & $-0.07$ & 0.46 \\
TGP       & -- & -- & -- & -- & $-0.09$ & 0.18 \\
TGP-giants& -- & -- & -- & -- & $-0.10$ & 0.23 \\
\hline               
\end{tabular}
\end{table}

\subsection{Atmospheric parameters and abundances determination}
\begin{figure}
   \centering
   \includegraphics[width=0.99\columnwidth]{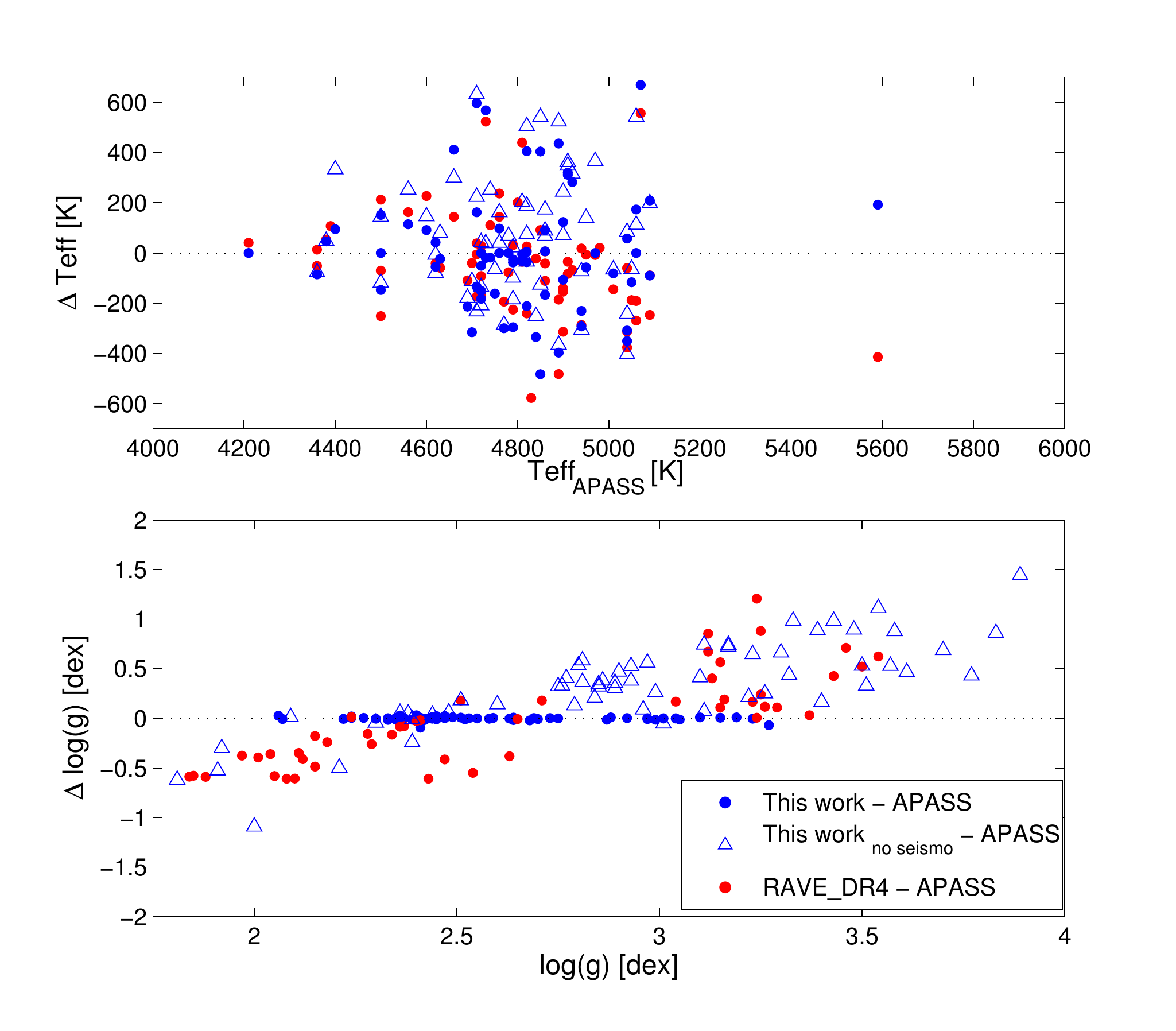}
   \caption{Top panel: comparison between the \Teff~derived in this work, with and without asteroseismology (blue points and blue triangles respectively) and in RAVE DR4 (red points) versus the one obtained using APASS photometry (Munari et al. 2014). Bottom panel: comparison between the \logg~derived in this work, with and without asteroseismology (blue points and blue triangles respectively) and in RAVE DR4 (red points) versus the one obtained using APASS photometry (Munari et al. 2014). }
              \label{Fig:phot}%
\end{figure}

For our analysis we considered the \Teff~and \logg$_{\rm seismo}$ derived using GAUFRE, adopting photometric \teff~as a prior and with the gravity fixed to the seismic \logg. We adopted then the \feh and individual element abundances derived by SP\_Ace, by fixing the \teff~and \logg~provided by GAUFRE. This strategy is needed since GAUFRE allows the iterative determination of the atmospheric parameters by using the photometric temperature as a prior (within an 600 K interval), while Sp\_Ace can take as input a fixed value of \teff~and \logg, and provides chemical abundances estimation. 

We determined the seismic \logg~using the \numax, the frequency corresponding to the maximum oscillation power. Starting from the scaling relation that links \numax to the stellar mass and radius (\citealp{Brown1991, Kallinger2010, Belkacem2011}):
\begin{equation}
\label{Eq:scMass}
{\nu_{\rm max} \over \nu_{{\rm max}, \, \odot}} = 
\left({M\over {\rm M}_\odot}\right) 
\left({R\over {\rm R}_\odot}\right)^{-2} 
\left({T_{\rm eff}\over {T}_{{\rm eff}, \, \odot}}\right)^{-1/2}
\end{equation}
It is possible to obtain a direct formula for the \logg~(by using the fundamental relation $g$=$G M/R^2$, where $G$ is the Newtonian gravity constant, $M$ is the stellar mass and $R$ is the stellar radius):
\begin{equation}
\label{Eq:sismologg}
\log g _{\rm seismo}= \log g_\odot + \log \left(
{{\nu_{\rm max} \over \nu_{{\rm max}, \, \odot}}}
\right)
+ {1 \over 2}
\log \left({T_{\rm eff}\over T_{{\rm eff}, \, \odot}}\right)
\end{equation}
with $\nu_{\rm{max}\odot}$ = 3140.0 $\mu$Hz \citep{Pinsonneault2014}, T$_{\rm{eff}\odot}$ = 5777 K, log(g)$_\odot$ = 4.44 dex.

This equation for the surface gravity is weakly sensitive to the effective temperature and, following that \numax can be well determined, it can provide \logg~with a precision better than 0.03 dex (\citealp{Kallinger2010,Morel2012}). For a discussion on the accuracy of the relation in Eq.~\ref{Eq:sismologg} see a discussion in  \cite{Davies2016}. Since the pipelines adopted in DR4 cannot work by fixing the log(g) to the seismic value, we performed our iterative analysis by using GAUFRE  and Sp\_Ace. 

Thanks to the tests discussed in Subsec.~\ref{Sec:Validation}, we determined the atmospheric parameters using the following strategy:
\begin{enumerate}
	\item Determination of the \logg$_{\rm seismo}$ adopting the APASS photometric temperature, using Eq.~\ref{Eq:sismologg};
	\item Analysis with GAUFRE by fixing the \logg~to the seismic value and using \teff$_{APASS}$ as prior (\Teff value can vary within a range of 500 K);
	\item Analysis with GAUFRE fixing the gravity to the \logg$_{\rm seismo}$ determined using the \teff measured at step 2
	\item Run GAUFRE iteratively until convergence (usually 3 iterations are needed);
	\item Run Sp\_Ace by fixing \logg~and \teff~to the values determined by GAUFRE for determining abundances. 
\end{enumerate}

\begin{figure}
   \centering
   \includegraphics[width=1\columnwidth]{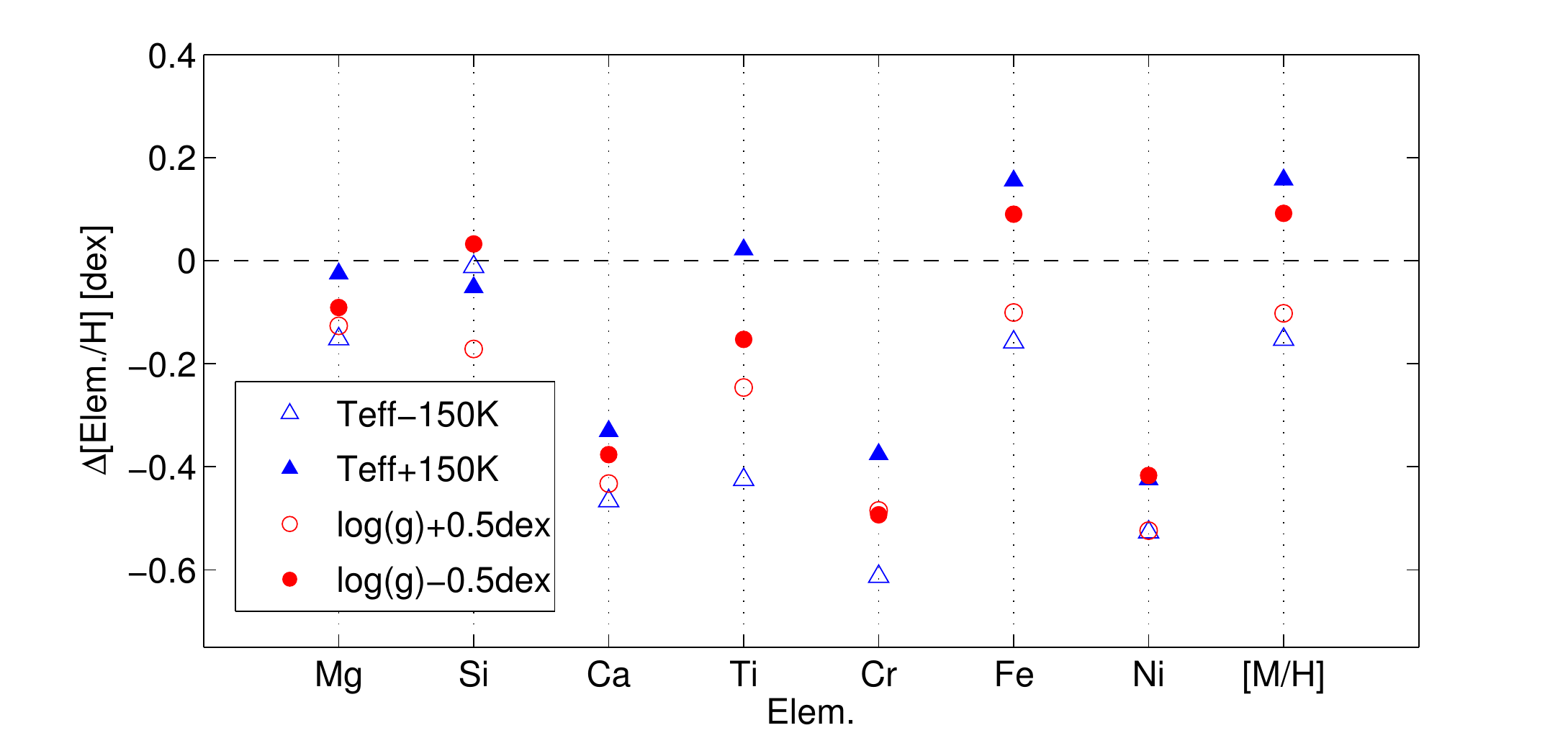}
   \caption{Element variations when applying a variation to \teff~and \logg~of $\pm$150 K and $\pm$0.50 dex. Differences between elements depend on the lines adopted and on the way their Curve of Growth (COG) depend on atmospheric parameters.}
              \label{Fig:elemVAR}%
\end{figure}

\begin{figure}
   \centering
   \includegraphics[width=1\columnwidth]{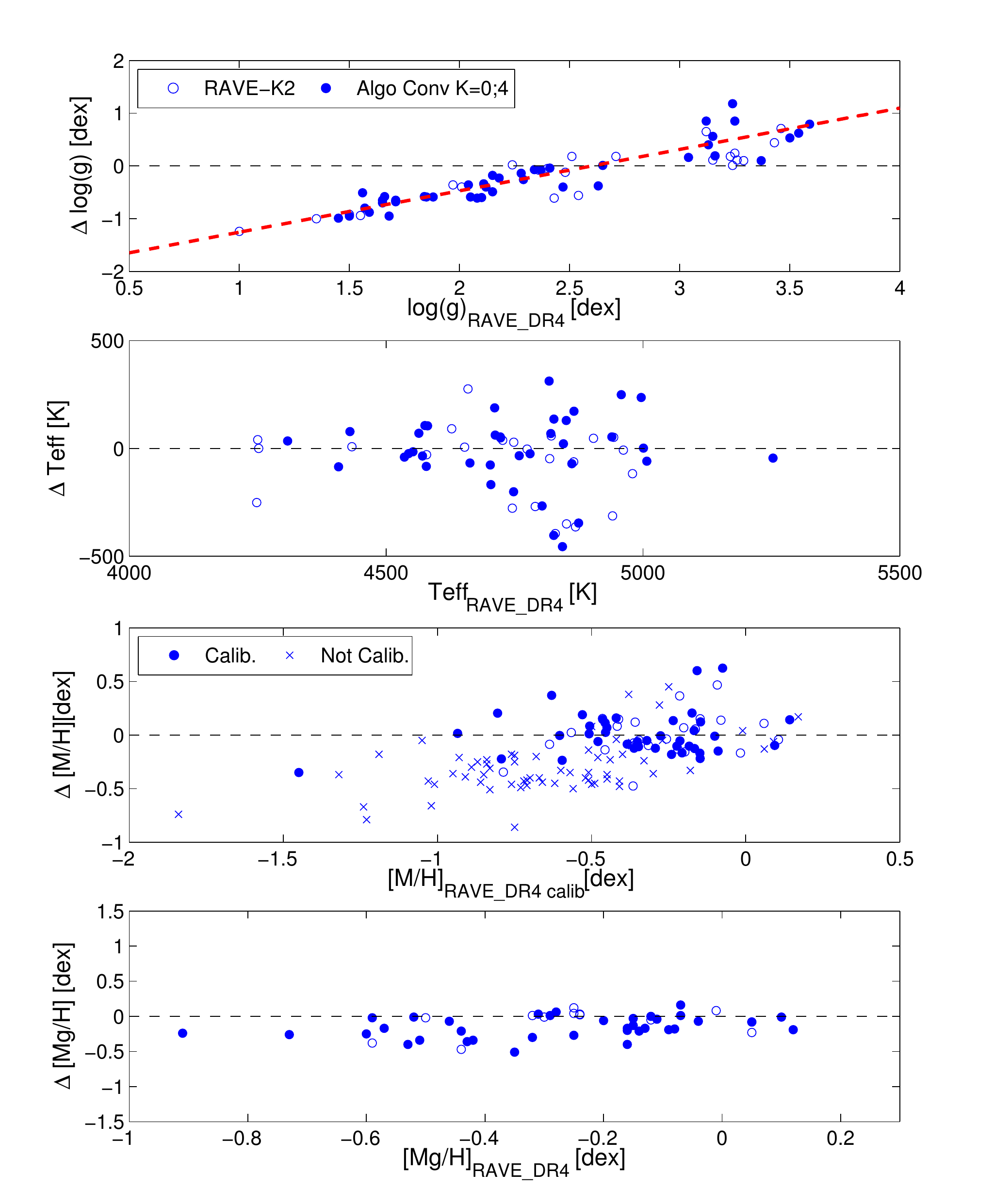}
   \caption{Difference in \logg, \teff, [M/H](calibrated and not calibrated) and [Mg/Fe] ($\Delta$ computed as RAVE DR4$-$this work) for the 62 RAVE targets where the GAUFRE+Sp\_Ace pipelines converged. On the top panel, the \logg~comparison, the fit used for calibrating \logg$_{\rm RAVE DR4}$ is shown (red dashed line).}
              \label{Fig:deltaparam}%
\end{figure}

The top panel of Fig.~\ref{Fig:phot} shows a comparison between the \citet{Munari2014} photometric \teff~with the temperatures derived in this work (with and without using asteroseismology) and those present in RAVE DR4. As the temperature increases, the dispersion of the difference in \teff~increases. This behaviour is partially due to the increase of the  differences in the reddening determination (hotter stars are intrinsically brighter and hence more distant than the colder stars at the same apparent magnitude). The bottom panel of Fig.~\ref{Fig:phot} shows the difference of the \logg~determined with different approaches (GAUFRE with no seismo, DR4 and GAUFRE with seismo) with respect to the \logg~computed using asteroseismology and the APASS photometric temperature. The strong degeneracy affecting the spectra causes the trend visible in the gravities determined from the pure spectroscopic analysis: the two pipelines show the same behaviour, even if using different approaches. 

\begin{table}
\label{Tab:errors}
\caption{Internal errors on atmospheric parameters and abundances in RAVE DR4 and those computed by combining spectroscopy and asteroseismology.}
\begin{tabular}{llcc} \hline \hline
$\sigma$         &  RAVE DR4 & This work\\ \hline
&  & \\ 
Teff [K] & 110 & 65 \\  
log(g) [dex]& 0.30 & 0.03 \\ 
$[\rm{Fe/H}]$ [dex]  & 0.10 & 0.08 \\  
$[\rm{elem./Fe}]$ [dex]& 0.20 & 0.08 \\ \hline
\end{tabular}
\end{table}

The method adopted in this work converged for 72 stars of the 87 analysed. The non-convergence of the pipelines was due to: bad SNR ratio (method not working for SNR<15), emission lines or non-corrected cosmic rays in the spectrum, metallicity too close (or outside) the pipeline's limits. The latter is the case of the two metal-poor stars; those stars are not present in this work, since their atmospheric parameters and abundances have been derived manually.

In Table~\ref{Tab:errors} are reported the typical internal errors on atmospheric parameters and abundances reported in the RAVE DR4 catalogue and those derived in this work. The adoption of the seismic \logg~improved significantly the accuracy of the \teff~and abundances measurement.

\subsection{Abundances measurement uncertainties}
\label{Subsec:elemVAR}

In order to understand the impact of strong offsets in temperature and metallicity to the element abundances determination, we re-derived the abundances of the benchmark stars using \Space by assuming the following shifts on stellar parameters: $\pm$ 150 K in \teff~and $\pm$ 0.5 dex in \logg. 

As expected, an overestimation/underestimation of \teff~and \logg~reflects into a overestimation/underestimation of the element abundance. The derived abundances of different elements vary differently, following the way the COG of the individual lines responds to the variation of \logg~and \Teff. 
Fig.~\ref{Fig:elemVAR} shows how the different elements (plus [M/H]) vary with respect to the value obtained by assuming the correct \teff~and \logg. 

\subsection{The RAVE spectra classification tool}

\begin{figure}
   \centering
   \includegraphics[width=1\columnwidth]{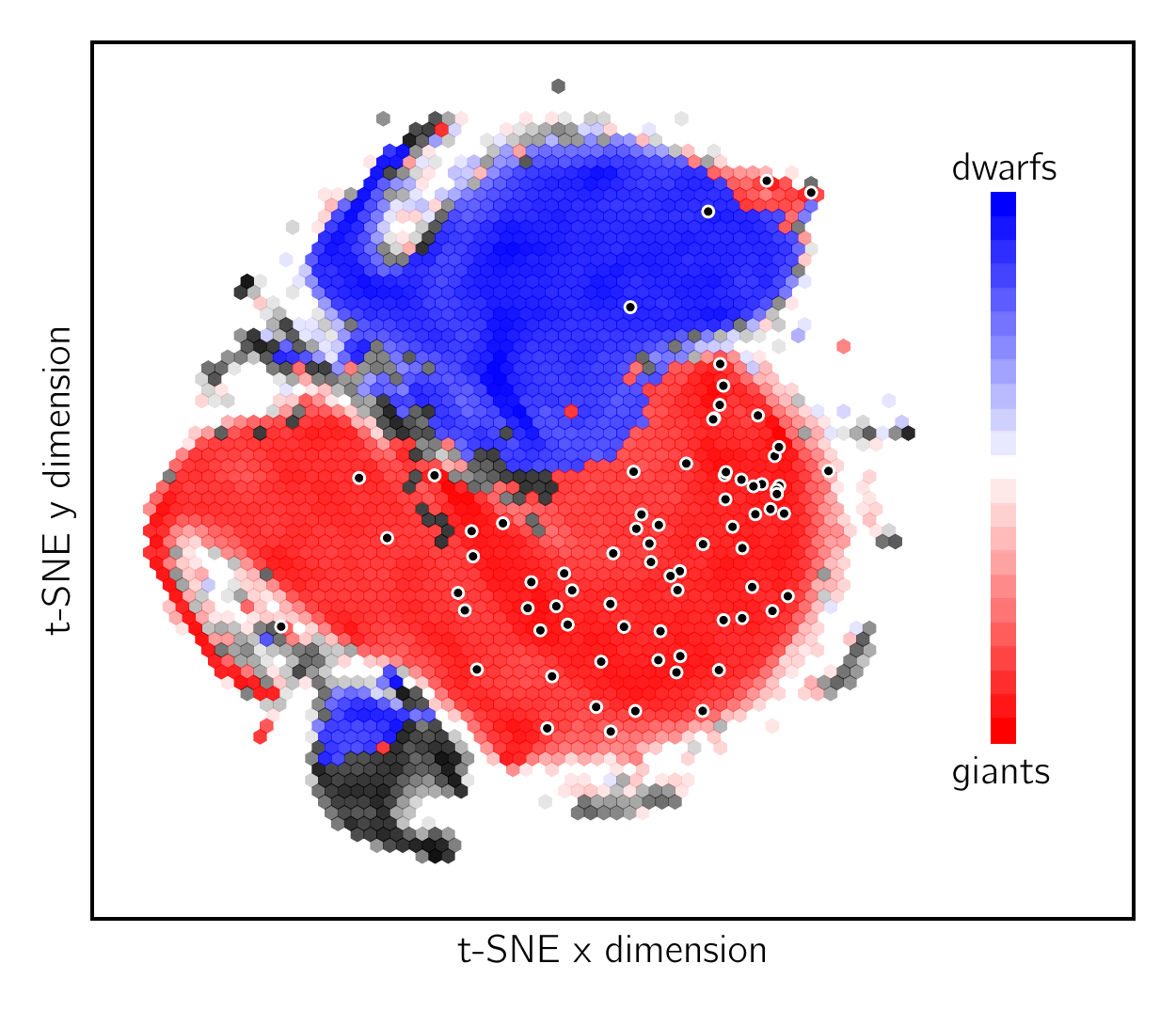}
   \caption{t-SNE projection of $\sim$420,000 RAVE spectra. The scaling in both direction is arbitrary,
therefore the units on the axes are omitted. The colour scale corresponds to the gravity of the
stars as computed by Kordopatis et al., 2013. Giants are shown in red and dwarfs in blue.
Lighter shaded hexagons include fewer stars than darker ones. Over-plotted black dots indicate
locations of the stars from this study.}
              \label{Fig:tSNE}%
\end{figure}

The diagram in Fig.~\ref{Fig:tSNE} shows the 2-dimensional t-SNE \citep{van2008visualizing}
projection of $\sim$420,000 RAVE spectra with SNR $>$ 10. Each spectrum was re-sampled to
768 common wavelength points and put into the data matrix that was used as an input to the t-SNE
dimensionality reduction method. The projection shown groups similar spectra together without
requiring any assumptions about the stellar parameters. Naturally, spectra of giant stars
being morphologically different from their dwarf counterparts are grouped in the different
parts of the projection. Besides the two main areas populated by the dwarfs and the giants,
the manifold also includes peninsulas and islands occupied by less regular types such as
spectroscopic binaries, hot stars, chromospherically active stars etc. It is obvious from the
figure that the majority of the stars from this study falls along the giant part of the manifold
($\log g<3.5$). There are two stars that fall onto the very metal-poor island (top right) and
two that reside in the dwarf region. The latter have very low SNR ratios therefore their
positioning in this diagram cannot be reliably used for confirmation of their gravity.

\section{Calibrating DR4 \logg: towards an improved DR5}
\begin{figure*}
   \centering
   \includegraphics[width=1\textwidth]{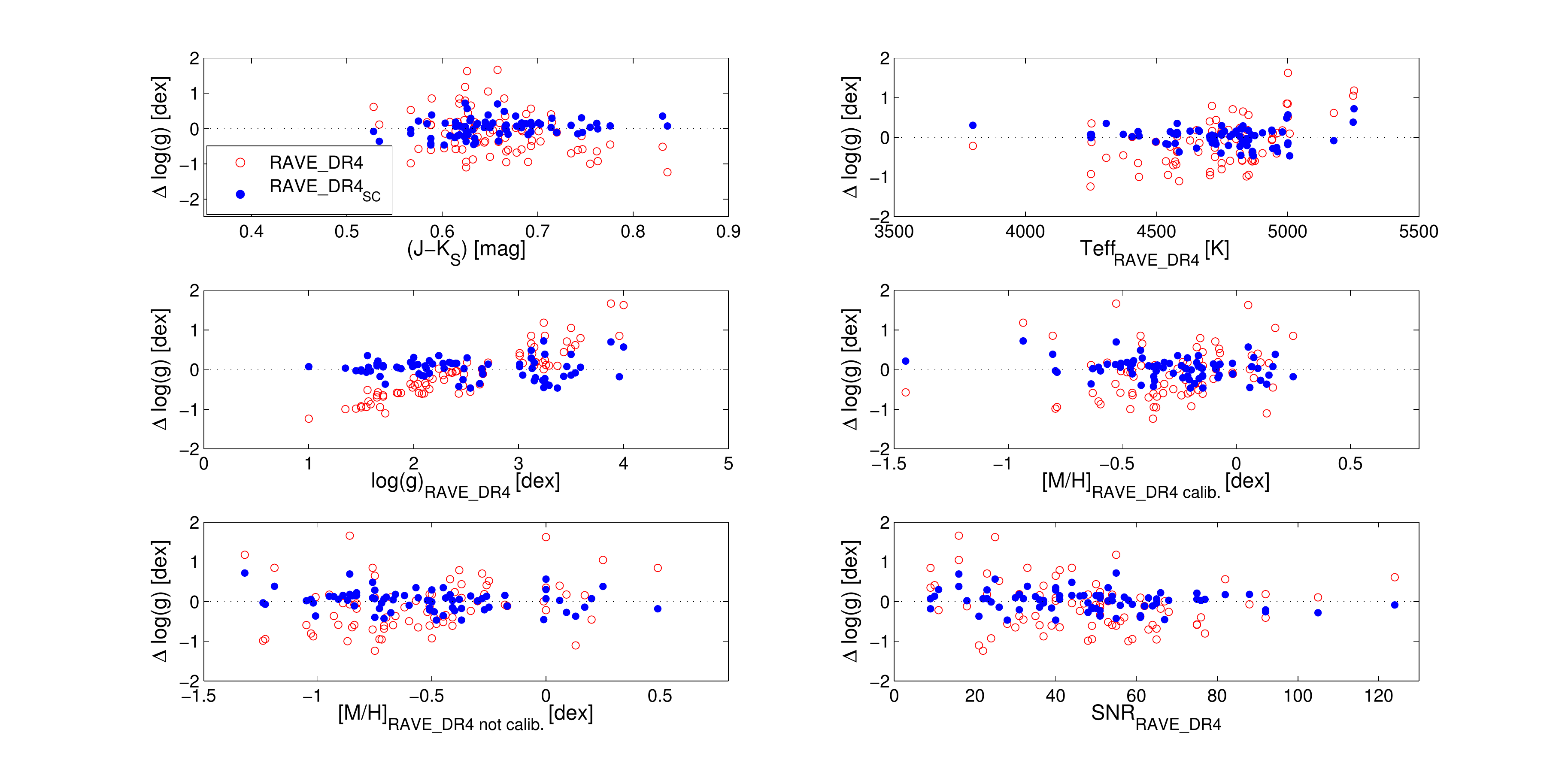}
   \caption{Difference in \logg~(computed as \logg$_{RAVE DR4}$-\logg$_{\rm seismo}$ - empty circles and \logg'$_{RAVE DR4}$-\logg$_{\rm seismo}$) vs (J-K$_S$) colour, \teff$_{RAVE DR4}$, \logg, calibrated and non-calibrated [M/H] and SNR. \logg' has been computing using Eq.~\ref{Eq:corrlogg}.}
              \label{Fig:calibration}%
\end{figure*}
RAVE red giants, and in particular red clump stars, had been widely used for investigating the properties of our Galaxy (e.g. \citealt{Bilir2012, Williams2013, Bienayme2014, Boeche2013a}). In these analyses giants and red clump stars were selected using photometric colour and a cut in \logg~(different approaches shown in Table~\ref{Tab:photsel}). Since all the stars used fall in the   0.5$\leq$(J-K$_S$)$\leq$0.8 and  1.8$\leq$\logg$_{RAVE}$$\leq$3.5 intervals, our sample of RAVE-K2 giants with asteroseismology is representative for understanding the offsets that can affect the RAVE red giants. In fact, RAVE-K2 red giants possess 0.5$\leq$(J-K$_S$)$\leq$0.8 and 1.3<\logg$_{RAVE}$<4 dex. 

As clearly visible in the top panel of Fig.~\ref{Fig:deltaparam}, there is a trend affecting DR4 \logg: the latest RAVE pipeline tends to distribute red giants on a wide gravity interval, classifying some giants as dwarfs or supergiants. This misclassification, as visible in Fig.~\ref{Fig:calibration}, does not depend on colour index, metallicity (calibrated or not-calibrated) or SNR. There is a trend of the \logg~depending on temperature, as one expects, since the two parameters are correlated.

   \begin{table}
      \caption[]{Selection criteria adopted in different works for creating the sample of red giants or red clump stars.}
         \label{Tab:photsel}
\centering                          
\begin{tabular}{l c c}       
\hline\hline                 
Work & Photometry & Spectroscopy\\  
\hline 
&  & \\                      
Bilir2012 (DR3)& (J-H)$_0$>0.4 & 2$\leq$\logg$\leq$3\\     
Williams2013 (DR3) & 0.55$\leq$(J-K$_S$)$\leq$0.8 & 1.8$\leq$\logg$\leq$3.0  \\
Bienayme2014 &  0.5$\leq$(J-K$_S$)$\leq$0.8 & 1.8$\leq$\logg$\leq$2.8   \\
Binney2015   & 0.5$\leq$(J-K$_S$)$\leq$0.8  & RC:1.7$\leq$\logg$\leq$2.4 \\
                     &                              &RG:2.4$\leq$\logg$\leq$3.5\\
Boeche2015   & --                           & 1.7$\leq$\logg$\leq$2.8 \\
                     &                              & 4250$\leq$\teff$\leq$5250 \\
Bovy2015     & 0.5$\leq$(J-K$_S$)$\leq$0.8  & --                        \\
\hline                                  
\end{tabular}
\end{table}

By using the K2C1-RAVE sample we obtained the following calibration for the RAVE DR4 gravities (plotted as dashed line in the top panel of Fig.~\ref{Fig:deltaparam}):

\begin{eqnarray}
\label{Eq:corrlogg}
\begin{split}
log(g)_{\rm RAVE\_DR~calib.}=log(g)_{\rm RAVE DR4}+\\-0.78^{0.88}_{0.68}\times(log(g)_{\rm RAVE DR4})+2.04^{1.78}_{2.29}
\end{split}
\end{eqnarray}

For the fit we considered only RAVE DR4 stars where the algorithm successfully converged (``Algo\_Conv\_K''=0 and ``Algo\_Conv\_K''=4, see \cite{Kordopatis2013} for the definition of these flags).

Since the difference in \logg~does not seem to depend on photometric colour, metallicity or SNR, the gravity calibration is only linearly depending on the original \logg$_{RAVE DR4}$.

The temperatures and abundances can be re-computed in order to obtain more consistent values for the RAVE giants.

\subsection{Sanity check: comparison with APOGEE and GES gravities}
\begin{figure}
   \centering
   \includegraphics[width=1\columnwidth]{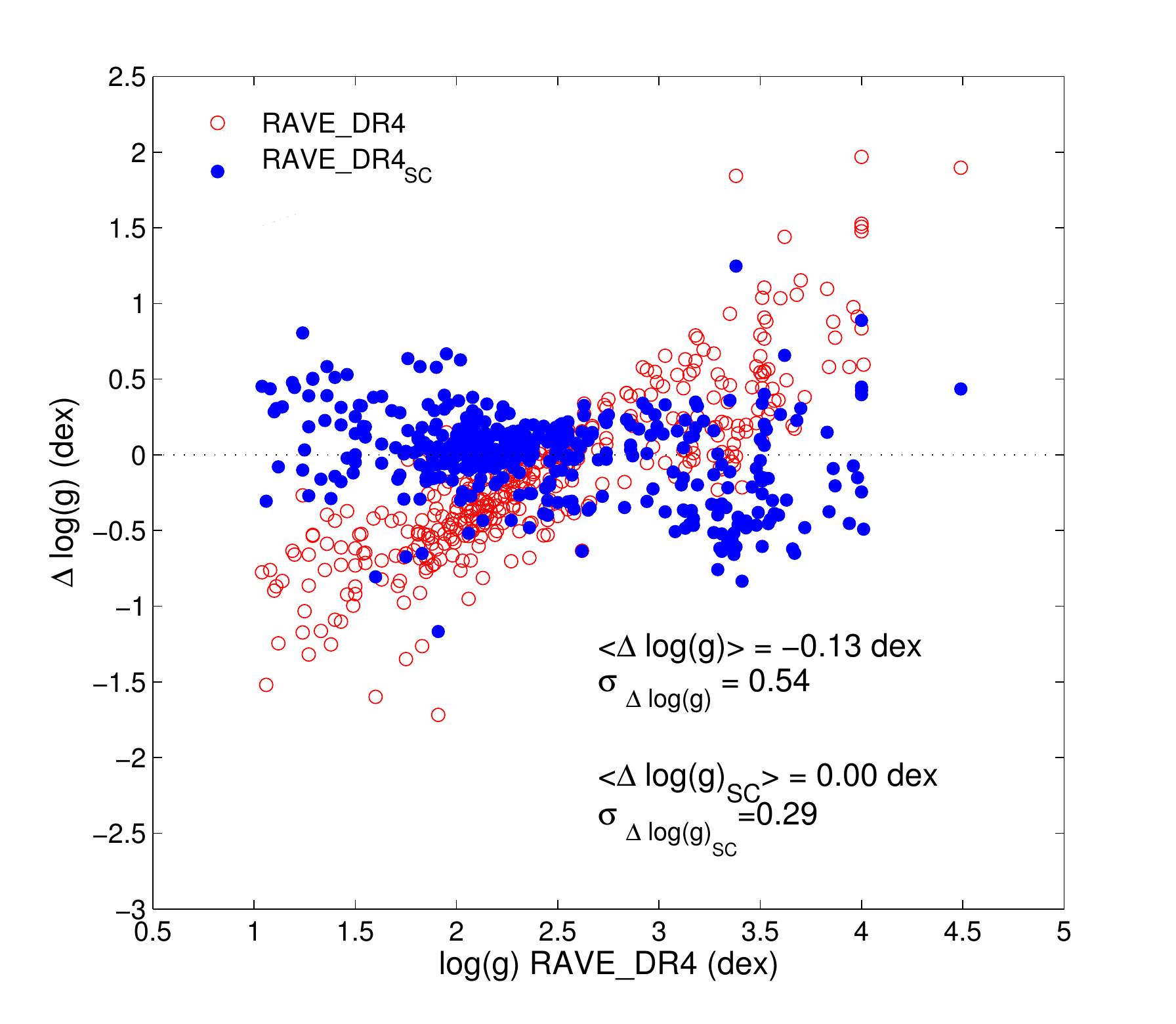}
   \caption{Difference in \logg~(computed as \logg$_{RAVE DR4}$-\logg$_{APOGEE}$) vs \logg$_{RAVE DR4}$ for the 855 RAVE targets in common with APOGEE-DR13.}
              \label{Fig:RAVE-APOGEE}%
\end{figure}
\begin{figure}
   \centering
   \includegraphics[width=1\columnwidth]{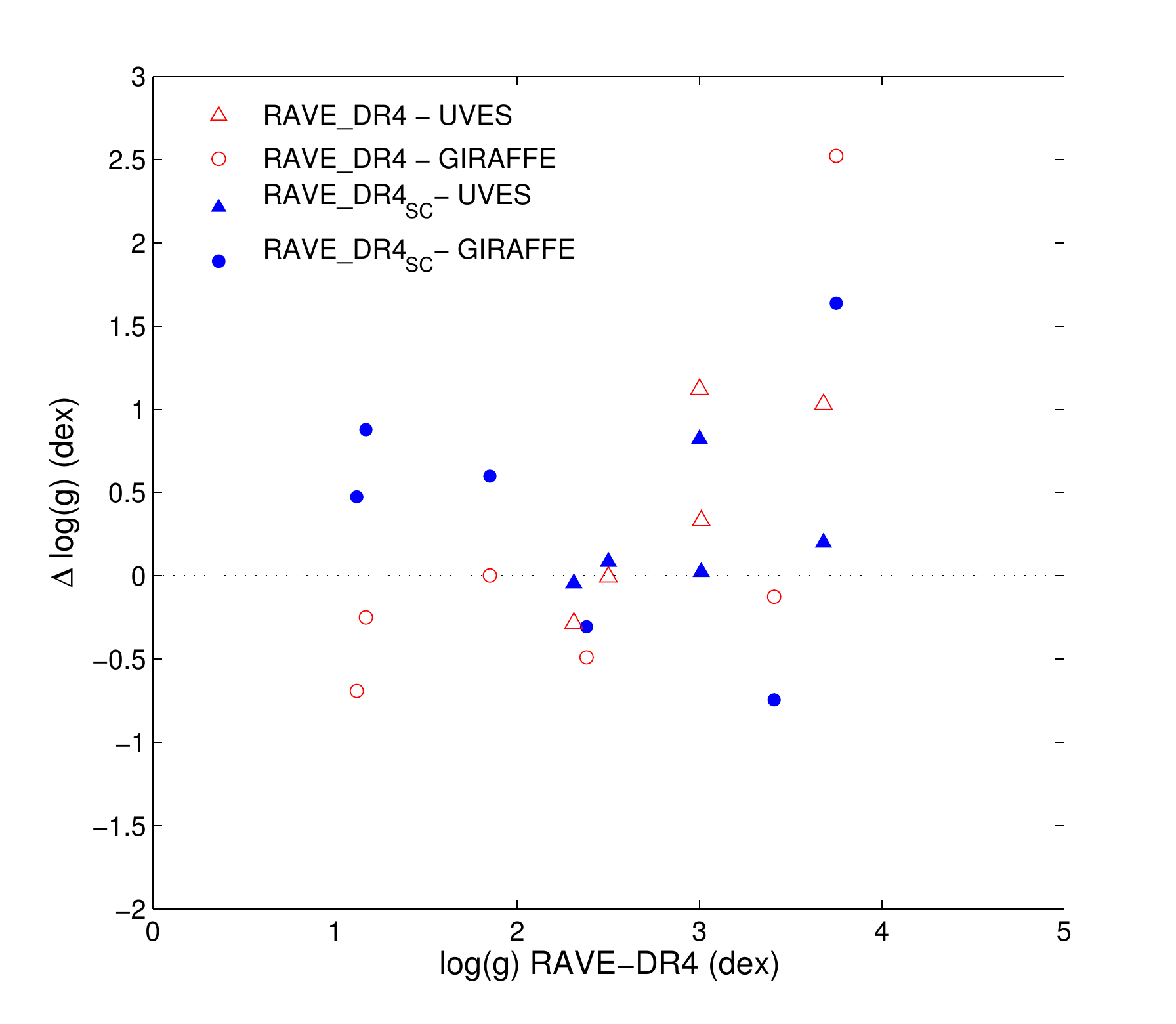}
   \caption{Difference in \logg~(computed as \logg$_{RAVE DR4}$-\logg$_{GES}$) vs \logg$_{RAVE DR4}$ for the 11 RAVE targets in common with GES. Triangles represent stars observed by GES using UVES, circles those observed using GIRAFFE instrument.}
              \label{Fig:RAVE-GES}%
\end{figure}

Since some RAVE red giants were observed by both APOGEE and GES surveys, we now compare  gravities of these targets with those present in RAVE DR4, and the new \logg s calibrated using Eq.~\ref{Eq:corrlogg}.\\

{\bf APOGEE}

There are 1422 targets in common between RAVE DR4 and APOGEE-DR13. Of those targets, 405 fulfil the quality criteria (convergence and quality flags for both RAVE and APOGEE) and lie in the colour interval $0.5<(J-K_S)<0.8$. A comparison between the \logg~is shown in Fig. \ref{Fig:RAVE-APOGEE}. The log(g) provided by APOGEE was calculated by applying an a-posteriori calibration to the \logg~measured by the pipeline. This APOGEE calibration was based on the seismic gravities of RGB stars from \kepler data, as described in \citet{Holtzman2015}. Since they used only the RGB stars for calibrating \logg, red clump gravities are overestimated by 0.2 dex. Figure \ref{Fig:RAVE-APOGEE} clearly shows that the  calibration adopted here for the RAVE DR4 gravities (see eq.~\ref{Eq:corrlogg}) leads to a good agreement with the APOGEE ones. Recently \citet{Hawkins2016} recomputed abundances for the APOGEE \kepler stars by fixing the gravity to the seismic one using the BACCHUS code, albeit without the iterative \Teff-\logg~ strategy adopted in this work.\\

{\bf GES}

There are 142 targets in common between RAVE DR4 and GES-DR4. Of those targets, 11 fulfil the quality criteria (convergence and quality flags for both RAVE and GES). A comparison between the \logg~is shown in Fig. \ref{Fig:RAVE-GES}. GES is providing homogenised atmospheric parameters and abundances, and in this work we considered F-G-K stars observed with UVES (high resolution, R=47,000) and F-G-K stars observed with GIRAFFE (low resolution, R=$\sim$19,000). The homogenisation is performed over the results provided by several pipelines (more than 10 nodes involved), and weighted following the performances of the several nodes on a calibration set of stars (GES paper in preparation). The consistency within the different approaches used by each node is guaranteed by the fact that all the nodes are using the same linelist (GES paper in preparation), same set of model atmospheres (MARCS, Gustafsson et al. 2008) and the same library of synthetic spectra (de Laverny et al. 2012). 

Although the number of red giants in common between the two surveys is not statistically meaningful, the trend is reduced when the correction of Eq.~\ref{Eq:corrlogg} is applied to RAVE \logg. In addition GES \logg~is the result of an homogenisation of different pipelines and it is not calibrated using asteroseismology.

\section{Impact of the adoption of the seismic \logg~on atmospheric parameters and abundances }

At medium resolution, the CaII triplet region is not providing very much information regarding stellar gravity, and this problem is present also in RAVE DR4.

By comparing the seismic \logg~with those provided in DR4, a clear trend is visible (first panel of Fig.~\ref{Fig:deltaparam}). The RAVE DR4 pipeline, in some cases, tends to identify giants as hot dwarfs or cold supergiants.
This misclassification is due to the \logg-\teff~degeneracies affecting the RAVE spectral interval. 

\subsection{Temperatures}

\begin{figure}
   \includegraphics[width=1\columnwidth]{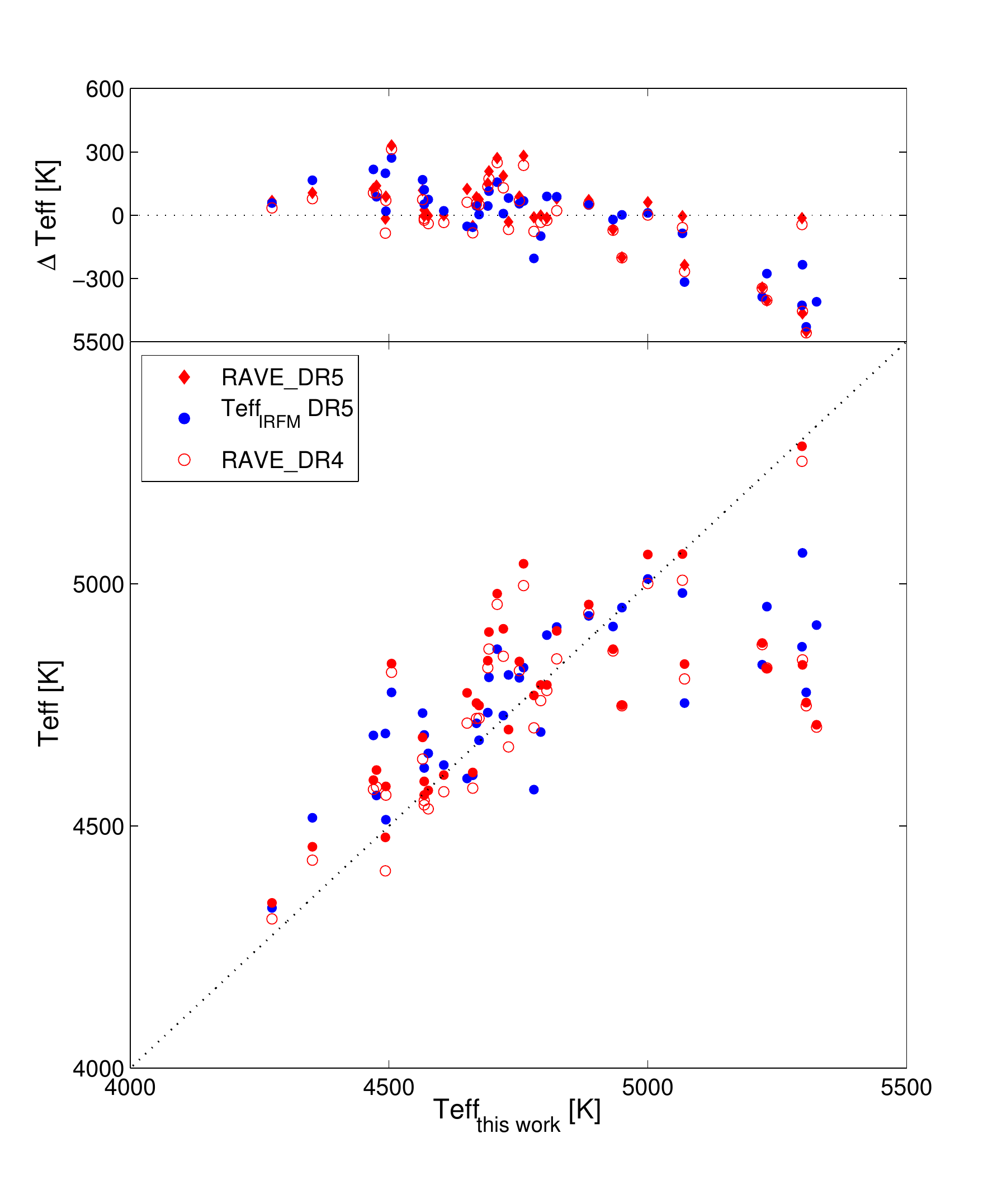}
   \caption{Effective Temperatures obtained in this work compared with those from the Infra Red Flux Methods (IRFM) temperatures (blue dots), RAVE(DR4) (open circles) and RAVE(DR5) (red points) .}
              \label{figtemp}%
\end{figure}

Figure~\ref{figtemp} shows a comparison of the temperature determined at the present work for our 72 RAVE-K2-C1 stars, with the values reported in DR4, DR5 and those of the IRFM as in DR5. A general agreement is found, except for the hotter stars. 

Indeed, it can be seen that at high temperatures (\Teff$>$5000 K) there is a discrepancy between the temperatures derived using \logg$_{seismo}$ and those present in DR5, the spectroscopic ones and those derived using the IRFM (adopting the method described in \citet{Casagrande2006}), and in DR4. As expected, the most deviant stars correspond to those for which there is a larger difference in \logg~respect to the seismic value (the top panel of  Fig.~\ref{Fig:diffP}). The discrepancy in \Teff~might be due to the \logg~discrepancy, since \citet{Casagrande2006} IRFM is slightly dependent to theoretical models, that for RAVE DR5 had been constructed using DR5 \Teff, \logg~and \feh. 

Thanks to the iterative process for deriving \logg~ and \Teff, we consider our temperatures reliable. The high precision of the \logg$_{\rm seismo}$ and the fact that it is weakly dependent to temperature, help in partially removing the degeneracy and in deriving an accurate temperature.

\subsection{Abundances}
\begin{figure}
   \includegraphics[width=1\columnwidth]{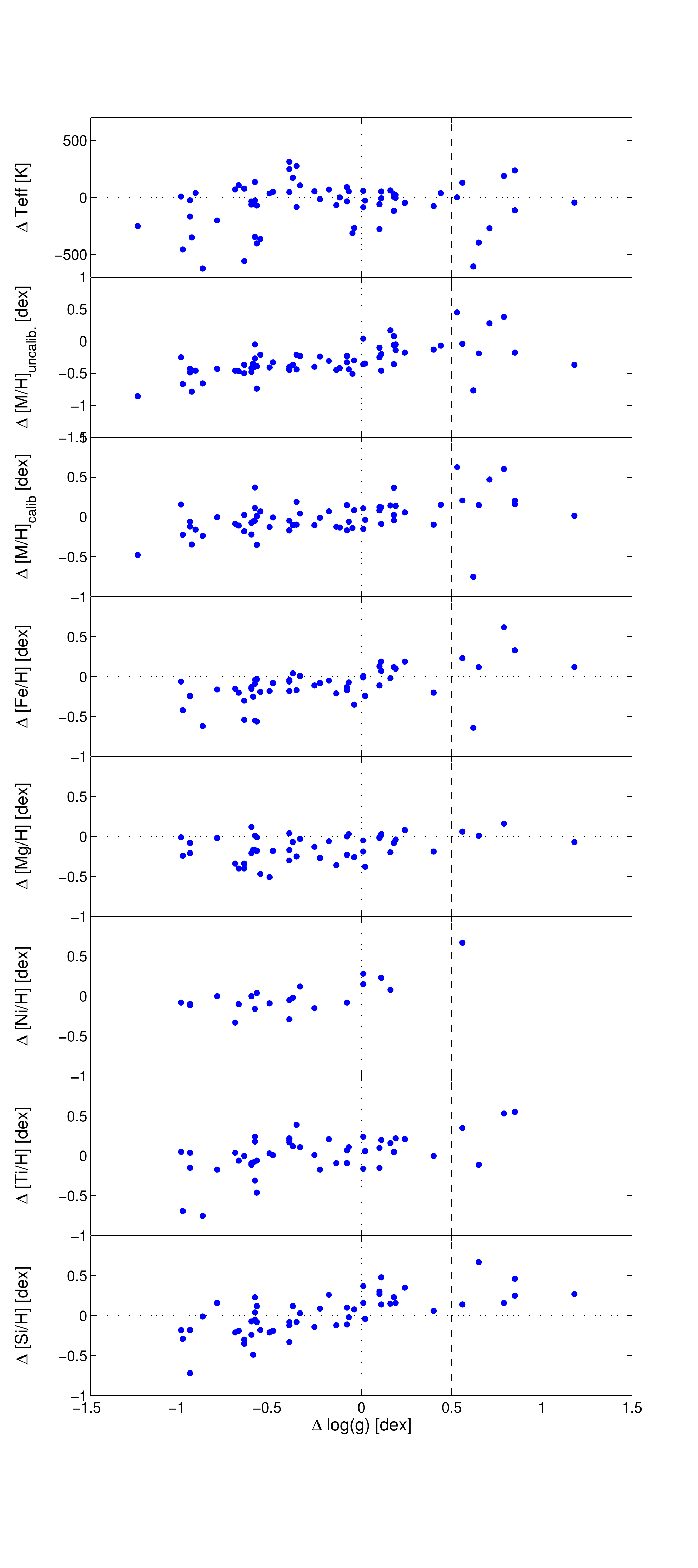}
   \caption{Differences in \teff, [M/H] (not calibrated and calibrated), \FeH, [Mg/H], [Ni/H], [Ti/H] and [Si/H] vs the differences in \logg. All differences are defined as RAVE(DR4)$-$(this work). The vertical dashed lines mark the $|\Delta log(g)|=$0.5 dex limits.}
              \label{Fig:diffP}%
\end{figure}
\begin{figure}
   \includegraphics[width=1\columnwidth]{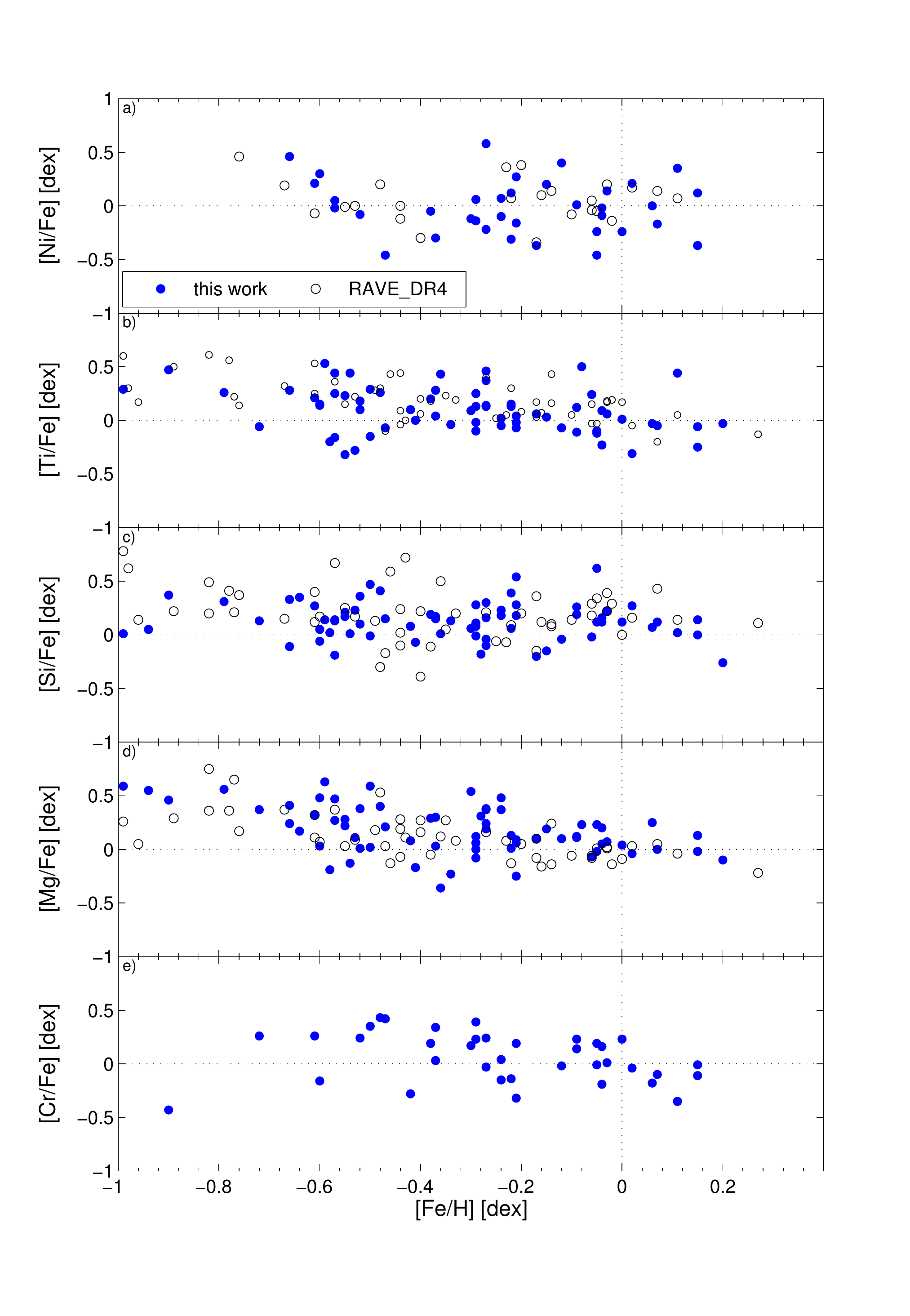}
   \caption{Distributions of alpha-elements (Ni,Ti,Si,Al,Mg) plus Cr versus Fe of the RAVE stars analysed in this work. Filled blue dots are the abundances obtained by using asteroseismology, black circles are the original DR4 values.}
              \label{Fig:elemFe}%
\end{figure}

The abundance determination is linked to the determination of the atmospheric parameters. Since \logg~and \teff~varied strongly from the spectroscopic determination of RAVE DR4 to the seismically determined one, we expect element abundances to vary as well.

Fig.~\ref{Fig:diffP} illustrates how the element abundances of Fe, Mg, Ni, Ti and Si (plus overall metallicity and temperature) vary depending on the difference in \logg. As expected, in general, when the DR4 gravity is underestimated, the element abundance is underestimated, and when the gravity is overestimated, the abundances are overestimated as well. The same happens for overall metallicity, both calibrated and not calibrated. However, the figure also shows that for the objects were the discrepancy between the gravities measured here and those of DR4 remains within 0.5 dex, the chemical abundances are only slightly affected.

Since the DR4 metallicity is calibrated following a function depending on \logg~and [M/H], there is an additional risk to introducing some metal-rich and metal-poor red giants as just the result of an erroneous \logg~determination plus an excessive metallicity correction.

The distributions of the $\alpha$-elements (Mg, Si and Ti) do not vary significantly with respect to DR4, as seen in panels $b$, $c$, $d$ of Fig.~\ref{Fig:elemFe}. The field is observing targets distributed perpendicularly to the Galactic plane (see Fig.~\ref{Fig:RZ_proj}), belonging to the thin and the thick disks. As one should expect for this field, Fe-poor objects are alpha-enhanced. Fe-peak elements (Ni and Cr, panels $a$ and $e$ of Fig.~\ref{Fig:elemFe}) do not vary following metallicity. Again, this trend follows what is expected, since Fe-peak elements are supposed to vary as Fe.

\subsection{DR5 calibration}

The results presented in this paper have been used to calibrate two catalogues in RAVE DR5: a) the main DR5 catalogue, which adopts a calibration for all stars (dwarfs and giants), computed using seismic \logg s from our 72 stars plus the Gaia benchmark stars, and b) the seismic calibrated catalogue of giants (DR5-SC) in the same colour range as the stars studied in this work, where the calibration adopted is the one presented in Eq.~\ref{Eq:corrlogg} (as in both the DR4 and DR5 the same spectroscopic pipeline is adopted). For the DR5-SC the chemical abundances were computed with calibrated gravities, and the IRFM temperatures (for a comparison of the CMD of the RAVE DR5 SC and RAVE DR5 catalogues, see Appendix). 

Figure~\ref{MDF} shows the metallicity distribution of the DR5 seismic calibrated catalogue in comparison with the MDF obtained for the same stars, but with DR5 (main catalogue) metallicities. Although similar, the DR5-SC MDF is narrower and has less metal-rich stars than the DR5 or DR4 MDFs. We also checked the MDF of the DR5-SC catalogue upon the removal of stars with temperatures above 5000 K (for which the IRFM temperatures differ from the ones obtained in our analysis, see Fig.~\ref{figtemp}), but the MDF did not change.

\begin{figure}
   \centering
   \includegraphics[width=1\columnwidth]{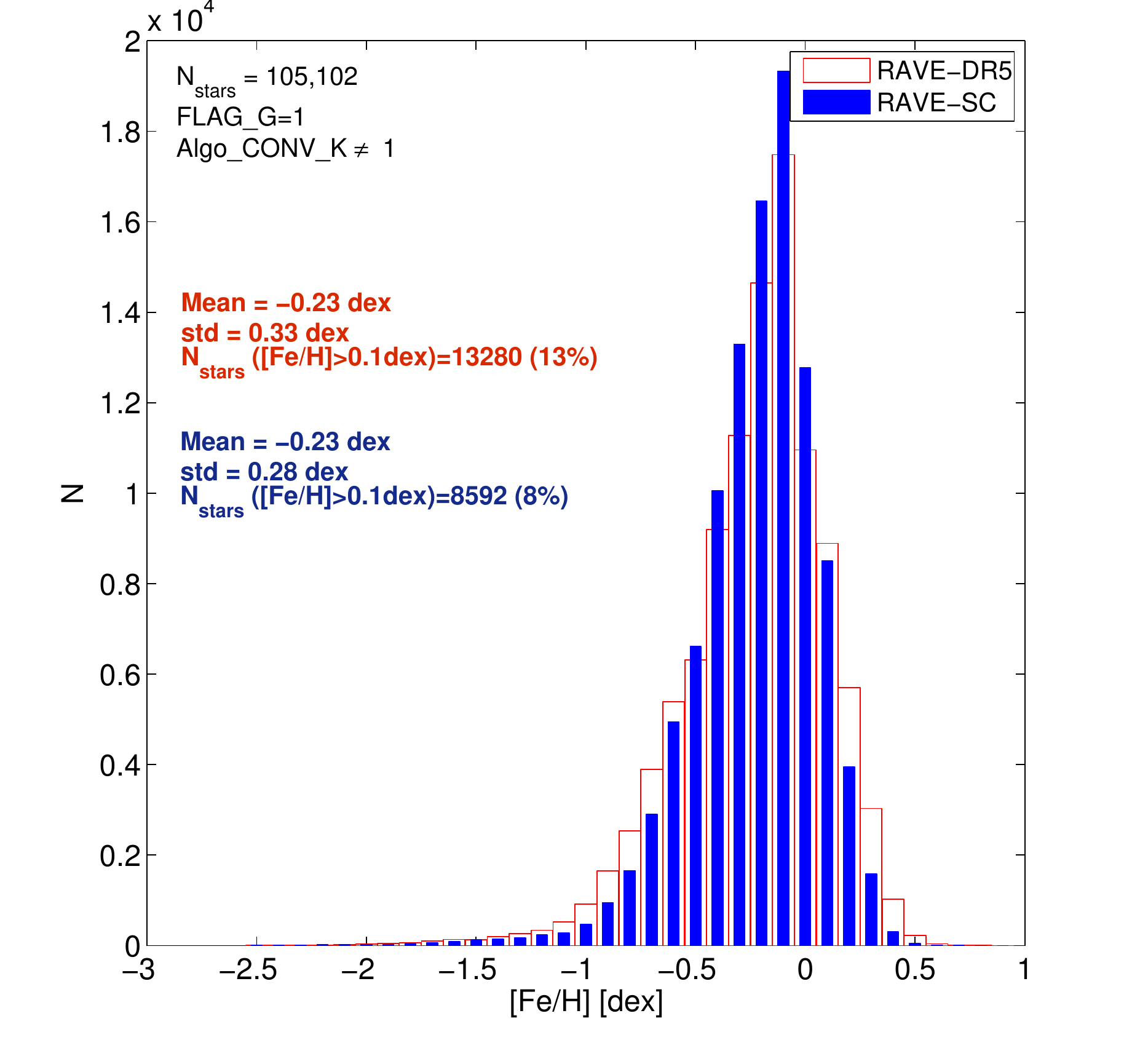}
   \caption{Metallicity distribution of RAVE DR5 seismic calibrated giants ( RAVE-SC, see \citet{Kunder2016}), compared with the MDF for the same stars but with DR5 metallicities}
              \label{MDF}
\end{figure}

\section{Distances, Reddening (and Ages)}

\begin{figure*}
   \centering
   \includegraphics[width=1\textwidth]{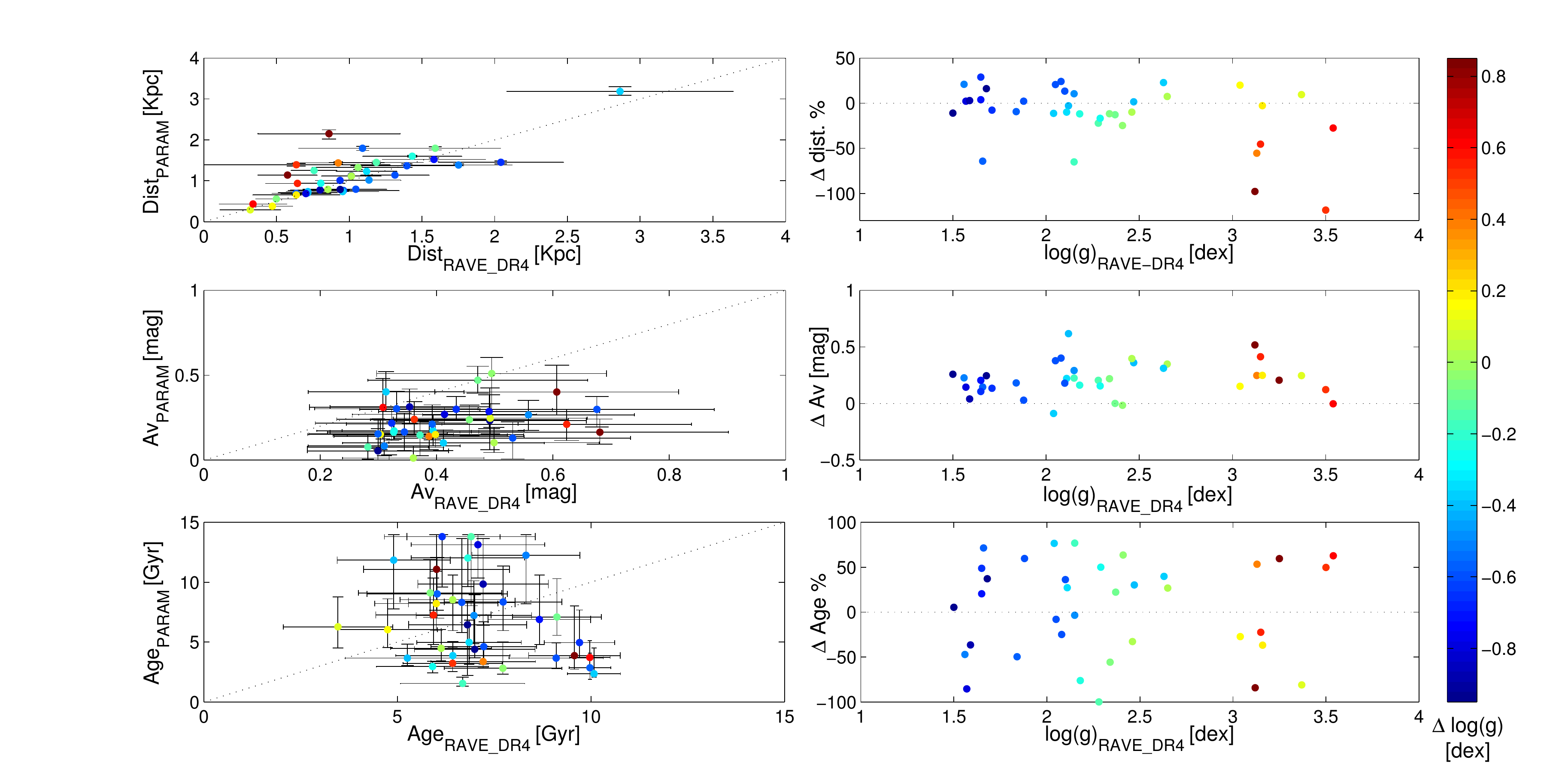}
   \caption{Comparison of distances, reddening (A$_V$) and age derived by using PARAM versus RAVE DR4 data. Data are colour coded following the difference in gravity. Differences for the various quantities are always computed as: $\Delta$=RAVE DR4 $-$ this work. 
Top left panel: distances derived by PARAM vs distance in RAVE DR4 (computed as the inverse of the parallax). 
Top right panel: distances residuals (in \%) vs \logg$_{\rm seismo}$. 
Left central panel: A$_V$ derived by PARAM vs A$_V$ present in RAVE DR4 catalogue. 
Right central panel: A$_V$ residuals vs \logg$_{\rm seismo}$. 
Bottom right panel: ages derived by PARAM vs RAVE DR4 ones. 
Bottom left panel: age residuals (in \%) vs \logg$_{\rm seismo}$.}
              \label{Fig:compPARAM}%
\end{figure*}

For our analysis we used masses, radii, distances, reddening and ages derived using the PARAM\footnote{http://stev.oapd.inaf.it/cgi-bin/param} tool (\citealp{daSilva2006, Rodrigues2014}), that derives stellar distance, reddening and age through Bayesian estimation. For this work we used the \citet{Rodrigues2014} version, implemented with the possibility of using seismic information ($\Delta\nu$, $\nu_{max}$ and evolutionary status). The code uses the seismic information by calculating \deltanu and \numax from the \citet{Bressan2012} set of isochrones using the scaling relations:

\begin{eqnarray}
\frac{M}{M_\odot} &\simeq& \left(\frac{\nu_{\rm max}}{\nu_{\rm max, \odot}}\right)^{3}\left(\frac{\Delta\nu}{\Delta\nu_{\odot}}\right)^{-4}\left(\frac{T_{\rm eff}}{T_{\rm eff, \odot}}\right)^{3/2} \label{eq:scalM}      \\
\frac{R}{R_\odot} &\simeq& \left(\frac{\nu_{\rm max}}{\nu_{\rm max, \odot}}\right)\left(\frac{\Delta\nu}{\Delta\nu_{\odot}}\right)^{-2}\left(\frac{T_{\rm eff}}{T_{\rm eff, \odot}}\right)^{1/2}\label{eq:scalR}
\label{eq:scalingrelations}
\end{eqnarray}
where $\nu_{\rm{max}\odot}$ = 3140.0 $\mu$Hz, \dnu$_{\rm{max}\odot}$ = 135.03 $\mu$Hz \citep{Pinsonneault2014}, T$_{\rm{eff}\odot}$ = 5777 K.

As input parameters we adopted the refined atmospheric parameters described in Section \ref{Sec:SpAnalys}, the seismic $\Delta\nu$ and $\nu_{max}$ described in Section~\ref{Sec:seismo}, and the photometric information from 2MASS, DENIS-$I$, AllWISE and APASS. PARAM converged for 67 stars (out of 72).

\begin{figure}
   \centering
   \includegraphics[width=1\columnwidth]{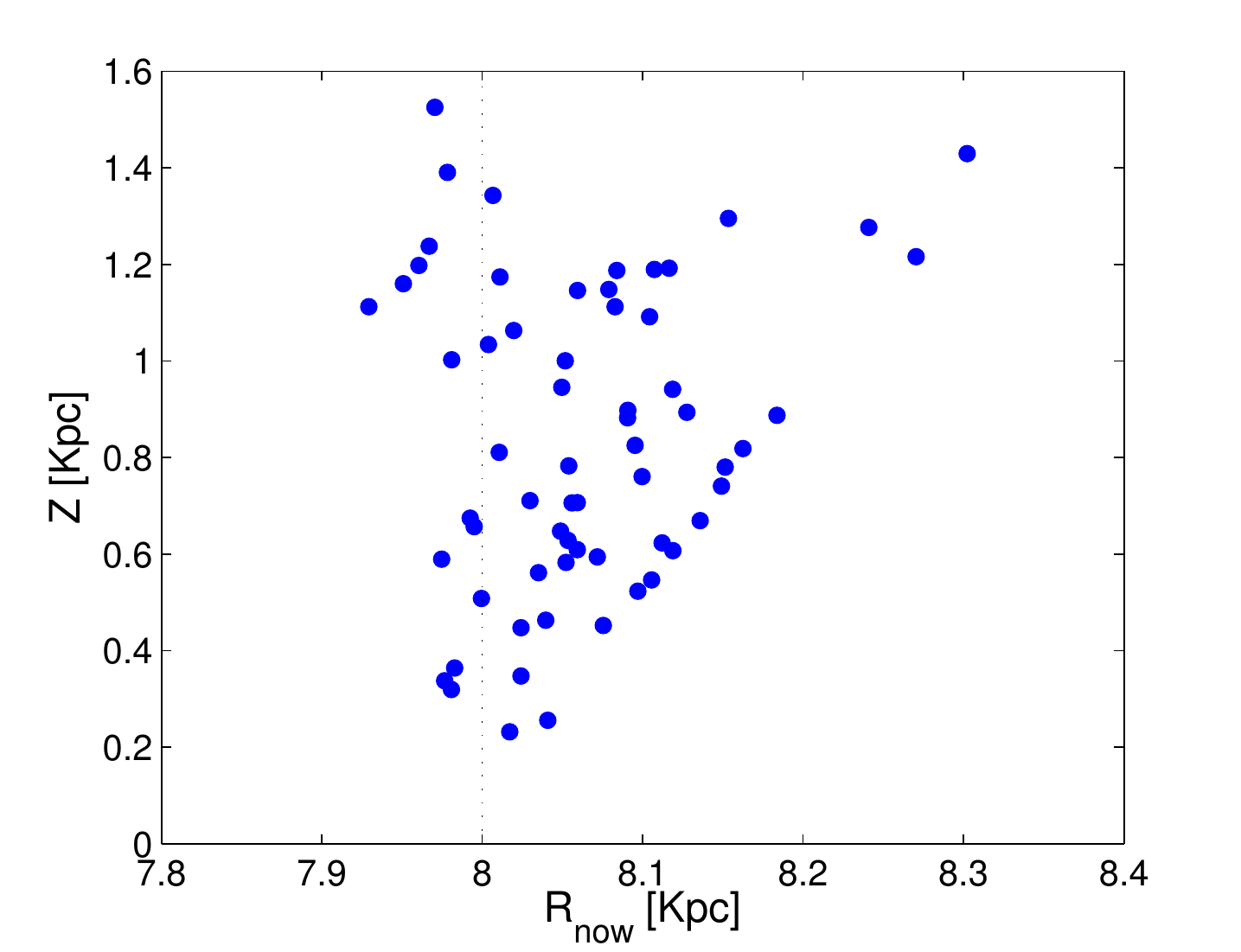}
   \caption{Distribution in Galactic Radius (R$_{now}$) and height to the Galactic plane (Z) of the RAVE targets, using the distances computed using PARAM.}
              \label{Fig:RZ_proj}%
\end{figure}
Fig.~\ref{Fig:RZ_proj} shows the spatial distribution, in Galactic Radius (R$_{now}$) and height to the Galactic plane (Z) of the stars. Stars are distributed perpendicularly to the Galactic plane, reaching a maximum Z of 1.5 kpc, with R$_{now}$ spanning from 7.9 to 8.3 kpc, and are thus representative of both the thick and thin disks.

Fig.~\ref{Fig:compPARAM} shows the comparison between distances, reddening (and ages) derived by PARAM, with those provided in RAVE DR4, distance and reddening offset and dispersion are reported in Table~\ref{Tab:mdsPARAMRAVE}. Since in this work we are not focusing on individual stellar ages and their individual errors, we consider the PARAM ages as a relative age indication, able to only discriminate old stars from intermediate and young objects. 

\begin{figure*}
   \centering
   \includegraphics[width=1\textwidth]{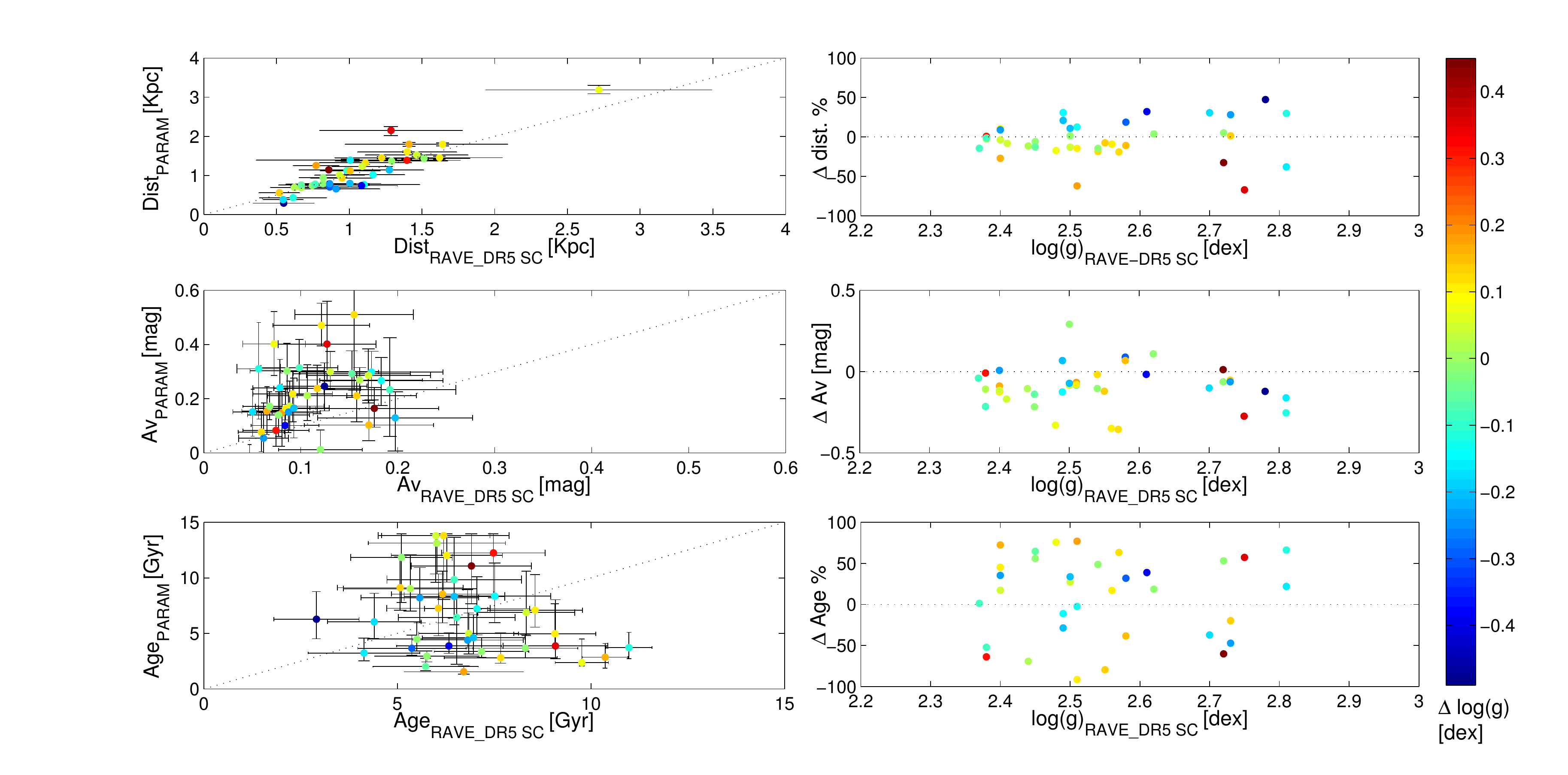}
   \caption{Same as Fig.~\ref{Fig:compPARAM}, but for RAVE DR5 seismic calibrated sample (see \citet{Kunder2016}) and flagged as FLAG\_G$=$1.}
              \label{Fig:compPARAM2}
\end{figure*}

Fig.~\ref{Fig:compPARAM2} is similar to Fig.~\ref{Fig:compPARAM}, but shows the comparison with the DR5-SC values (results of the comparison with RAVE DR5 main catalogue is not shown, as the results are similar to the ones shown here). 

   \begin{table}
      \caption[]{Means and dispersions of the difference between RAVE DR5 (general catalogue and seismic calibrated) and PARAM distances and reddening.}
         \label{Tab:mdsPARAMRAVE}
\centering                         
\begin{tabular}{l c c}      
\hline\hline

             & \multicolumn{2}{l}{RAVE\_DR5}\\ \hline   
  & Distance & Reddening \\
  & & [mag]  \\ 
                     
$\Delta$ & 21\% & $-$0.08  \\   
$\sigma$ & 48\% & 0.13  \\

\hline  

& \multicolumn{2}{l}{RAVE\_DR5 SC}\\ \hline 

  & Distance & Reddening\\
  & & [mag] \\                  
$\Delta$ & 3\% & $-$0.09  \\    
$\sigma$ & 23\% & 0.12  \\
\hline  

& \multicolumn{2}{l}{RAVE\_DR4}\\ \hline 

  & Distance & Reddening\\
  & & [mag]  \\                 

$\Delta$ & 14\% & $-$0.20  \\    
$\sigma$ & 34\% & 0.14 \\

\hline

\end{tabular}
\end{table}

In the distances comparison with RAVE DR4, we considered distances derived from parallaxes, as suggested by \cite{Binney2013}. Red clump gravities in DR4 are overestimated by $\sim$0.3 dex, leading to a distance overestimation  of $\sim$25\%. The same problem can happen also with the rest of the red giants. The adoption of a imprecise \logg~and reddening, results in an overestimation or underestimation of the distance. An object with an overestimated gravity is less bright, and therefore it appears closer (the contrary happens when the \logg~is underestimated). This behaviour is visible in the top row of Fig.~\ref{Fig:compPARAM}. 

The differences in gravity and distance impact also the derived reddening. An object that in DR4 possesses a \logg~in agreement with the seismic values, but has a lower distance, possesses a Av that is underestimated. And the opposite behaviour happens when the object has a larger distance than the one determined in this work (see middle panels of Fig.~\ref{Fig:compPARAM}). In addition, reddening in RAVE DR4 is systematically overestimated by 0.20 mag respect to the reddening derived using PARAM (see also Table~\ref{Tab:mdsPARAMRAVE}). 

As explained in \citet{Kordopatis2013} ages in DR4 are only indicative, since in the Bayesian computation of the distance (and hence mass and age), stars were assumed as ``old''. As visible in the bottom panels of Fig.~\ref{Fig:compPARAM}, ages computed using PARAM, instead, show that the RAVE-K2 Campaign 1 stars cover a wider age interval, from 1 to 13.7 Gyrs. 

Figure~\ref{Fig:compPARAM2} shows instead a comparison with the results of DR5-SC catalogue, which shows a slight improvement thanks to the combination of photometric and seismic information.

\begin{figure}
  \centering
  \includegraphics[width=1\columnwidth]{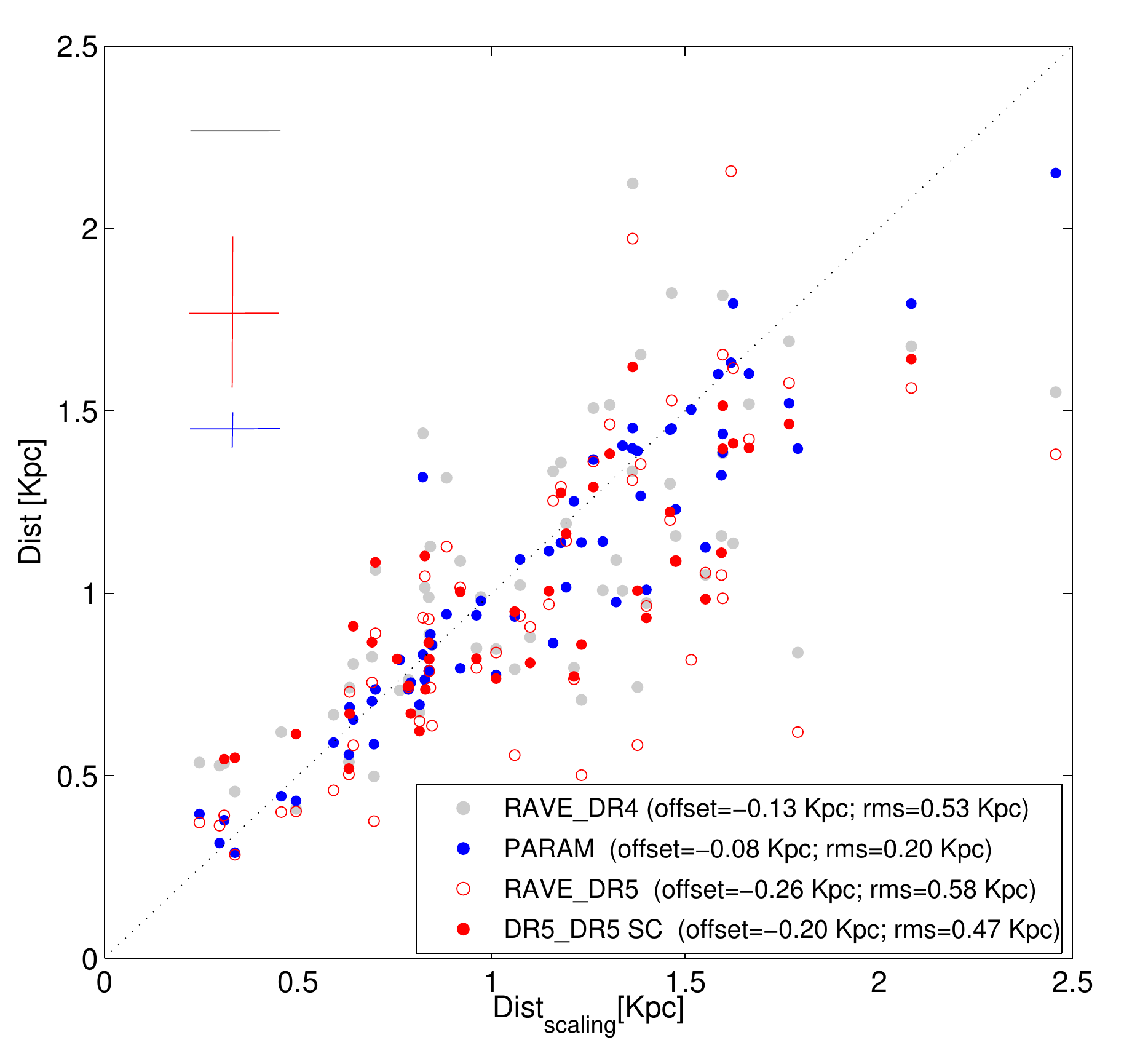}
   \caption{Comparison of the distances obtained using asteroseismology and the direct method adopted in \citep{Miglio2013}, with the distances provided in RAVE (DR4, DR5 and DR5-SC) and the distances determined using PARAM (blue points).  Typical errors of each method are shown in the top-left of the figure (both DR5 and DR5-SC have the same typical error).}
              \label{Fig:compSchlSeism}%
\end{figure}

Finally, we also show the distances computed by using only asteroseismology and magnitude, using the direct method described in \citet{Miglio2013}:

\begin{eqnarray}
\begin{split}
\log{d}=1+2.5\log{\frac{T_{\rm eff}}{T_{\rm eff,\odot}}}+\log{\frac{\nu_{\rm max}}{\nu_{\rm max,\odot}}}+\\-2\log{\frac{\Delta\nu}{\Delta\nu_\odot}}+0.2(m_{\rm V}+BC_{\rm V}-A_{\rm V}-M_{\rm bol, \odot})
\end{split}
\label{eq:seismodistance}
\end{eqnarray}
where the solar values are the same adopted in Eq.~\ref{eq:scalingrelations}, the Landolt V magnitude comes from APASS catalogue, A$_{\rm V}$ is the Schlegel reddening, and the bolometric correction (BC) is taken from \citet{Girardi2002}. {\bf The error on the distance determined using Eq.~\ref{eq:seismodistance} was computed using propagation of uncertainty. The median uncertainty is of 10\%, by taking into account the errors on \dnu and \numax, temperature, magnitude and reddening.} A comparison of the direct distances with distances provided by RAVE (DR4, DR5, and DR5 SC) and those computed in this work (PARAM) are shown in the bottom panel of Fig.~\ref{Fig:compSchlSeism}. The distances computed using PARAM show good agreement with those computed with the direct method, while a larger dispersion is present in the RAVE DR4 distances, likely as a consequence of the different atmospheric parameters  and their larger errors adopted and of the use of seismic information by PARAM.  The typical error on distance of the previously mentioned methods is of 25\% for DR4, 24\% in DR5, 24\% in DR5-SC and 4\% in PARAM.
 
\section{Conclusions}

In this paper we analysed  87 RAVE stars with detected solar like oscillations, observed during Campaign 1 of the K2 mission. The use of
asteroseismic \logg~ (with typical accuracy of 0.03 dex), and photometric temperature, was able to break  the  \logg-\teff~degeneracy that affects the RAVE wavelength interval (around CaII Triplet, especially for red giants). By comparing our measurements with those of  RAVE DR4, we were able to quantify the impact of the refined gravities and effective temperature obtained here on the elemental abundances, distances, and reddening (and age) determinations for these stars.\\
 
Our results can be summarized as follows:
\begin{itemize}
\item A difference between \logg$_{\rm seismo}$ and \logg$_{\rm RAVE~DR4}$ exists. This is a consequence of the resolution and short spectral coverage of the RAVE survey, that leads to a strong \logg-\teff~degeneracy. This degeneracy had been partially solved in RAVE DR4 by adopting a decision-tree pipeline, together with a projection-method one. In this work we  provide a calibration for the gravity of RAVE DR4 red giants (Eq.~\ref{Eq:corrlogg}) that is valid for giants selected in the colour interval 0.50$\leq$(J-K$_S$)$\leq$0.85. 
  
\item The difference in \logg~leads, as expected, to differences respect to the newly recomputed \teff, overall metallicity [M/H], and single element abundances. Stars with an overestimated gravity in DR4, have overestimated \teff~and metallicity.

\item The change of the \logg~leads to a change of the star's luminosity, affecting distances and reddening. A correct sample of red giants, with distances in agreement with the distances derived in this work, can be selected from RAVE DR4 by applying a colour cut  0.50$\leq$(J-K$_S$)$\leq$0.85 and a very narrow cut in \logg,  2.5$\leq$\logg$\leq$2.8 dex.
 
\end{itemize} 

We determined a calibration for \logg~following Eq.~\ref{Eq:corrlogg}, for photometrically selected giants in DR4. The same correction was used for the red giants in the forthcoming RAVE Data Release (DR5). In the RAVE DR5 catalogue seismically calibrated gravities were provided for a sample of red giants, photometrically selected using 0.50$\leq$(J-K$_{S}$)$_0$$\leq$0.85. These gravities appear in the ``LOGG\_SC'' column. We also recommend to recompute abundances, metallicity, abundances and distances using the calibrated \logg. The shifts introduced by a uncertain \logg~assumption may introduce artefacts, such as metal-rich or metal-poor stars, or stars with the incorrect distance or kinematics. In the RAVE DR5 catalogue this re-computation has been already performed.

The nature of these trends will be further explored in the other K2 Campaigns, increasing the statistics of our calibration sample and using RAVE stars possessing asteroseismology for Galactic Archaeology investigations. Gaia will help improving the atmospheric parameters as well. The strategy developed in this work can be used for the future parameter determination, by using the \teff~and the \logg~coming from independent sources as priors (e.g. magnitude colours, parallaxes).

\begin{acknowledgements}
AM, WJC, GRD, and YPE acknowledge the support of the UK Science and Technology Facilities Council (STFC). TSR acknowledges support from CNPq-Brazil. This work has made use of the VALD database, operated at Uppsala University, the Institute of Astronomy RAS in Moscow, and the University of Vienna. Funding for RAVE has been provided by: the Australian Astronomical Observatory; the Leibniz-Institut fuer Astrophysik Potsdam (AIP); the Australian National University; the Australian Research Council; the French National Research Agency; the German Research Foundation (SPP 1177 and SFB 881); the European Research Council (ERC-StG 240271 Galactica); the Istituto Nazionale di Astrofisica at Padova; The Johns Hopkins University; the National Science Foundation of the USA (AST-0908326); the W. M. Keck foundation; the Macquarie University; the Netherlands Research School for Astronomy; the Natural Sciences and Engineering Research Council of Canada; the Slovenian Research Agency; the Swiss National Science Foundation; the Science \& Technology Facilities Council of the UK; Opticon; Strasbourg Observatory; and the Universities of Groningen, Heidelberg and Sydney. 
The RAVE web site is at: \url{https://www.rave-survey.org}.
\end{acknowledgements}

\begin{appendix}
\section{The $\chi^2$ fitting with a library of synthetic spectra}

There are several causes to the rigid shift in metallicity pointed out in Section 4.2. It can be caused by wrong continuum placement and by degeneracies, but also by a wrong assumption of the line parameters (e.g. log$gf$).

\begin{figure*}
   \centering
   \includegraphics[width=1\textwidth]{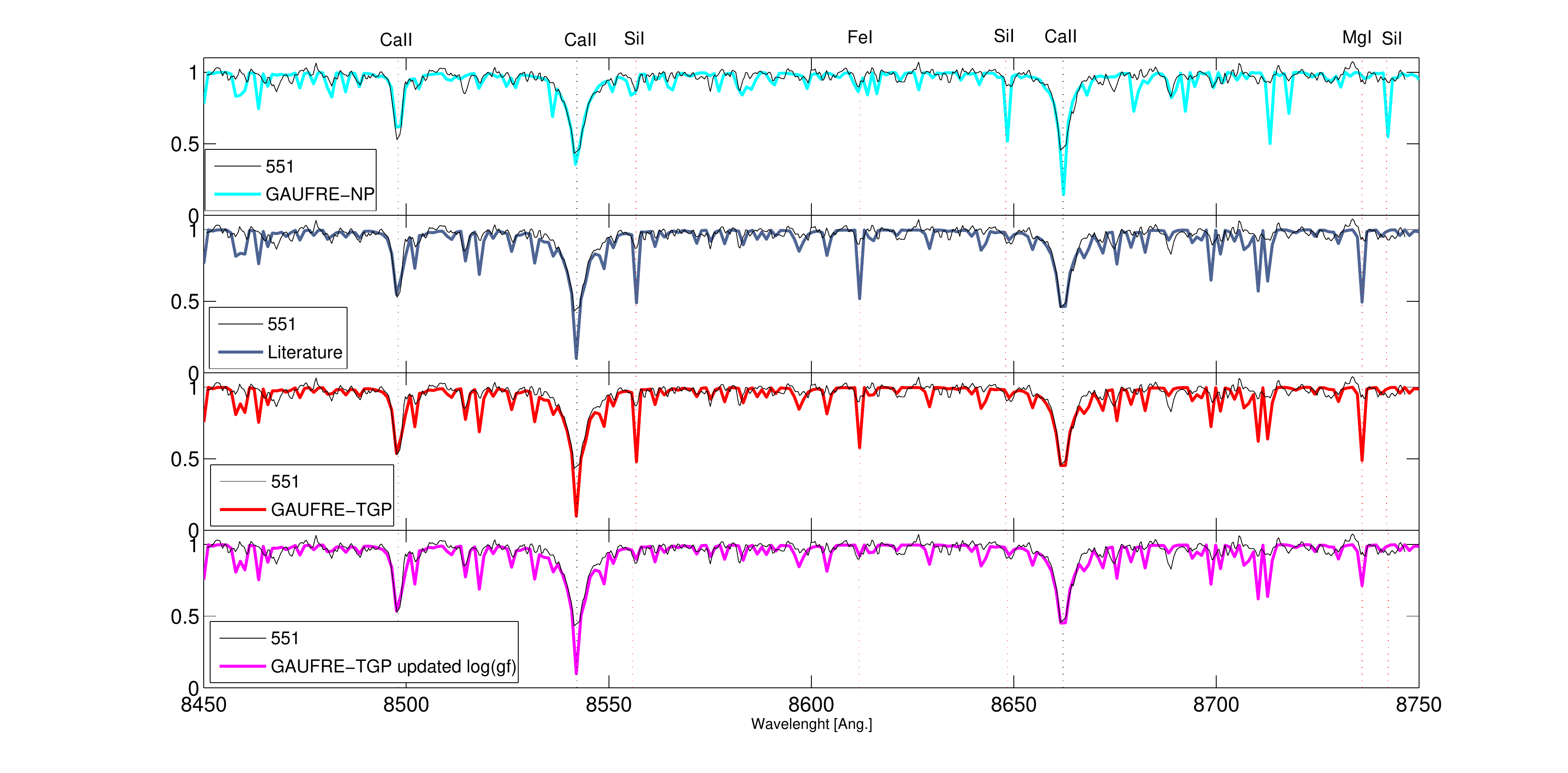}
    \caption{Comparison of the RAVE observed spectrum with a synthesized one, for the benchmark star 551 . From top to bottom: first panel, comparison of the real spectrum with the synthetic one, computed using the GAUFRE-NP parameters. Second panel: comparison of the real spectrum with the synthetic one computed using the literature parameters. Third panel: comparison of the real spectrum with the synthetic one built using the GAUFRE-TGP parameters. Bottom panel: comparison of the real spectrum with the synthetic one computed using the GAUFRE-GTP parameters and the log$gf$ adopted in SP\_Ace.}
              \label{Fig:specSynth}%
\end{figure*}

We identified some of these lines with wrong log$gf$, as visible in Fig.~\ref{Fig:specSynth} and summarized in Table~\ref{Tab:lines}. 

   \begin{table*}
      \caption[]{Example of lines that possess an incorrect log$gf$ values in the VALD3 linelist and their log$gf$ as in \Space linelist.}
         \label{Tab:lines}
\centering                          
\begin{tabular}{l c c c c}       
\hline\hline                 
Species & Wavelength & \multicolumn{2}{c}{VALD3} & SPACE\\
        &            & log$gf$ & source & log$gf$\\  
\hline 
&  & & & \\
SiI &  8555.903 \AA & $-$3.127 & K07& $-$2.39\\
SiI &  8556.777 \AA & $-$0.151 & K07& $-$0.35\\
SiI &  8555.805\AA & $-$0.407 & K07& $-$0.55\\                      
FeI &  8610.610 \AA & $-$2.683& K14 & $-$1.76\\
FeI &  8611.803 \AA & $-$1.926& K14 & $-$2.00\\
MgI &  8736.02 \AA & blended multiplet & NIST10 & -0.26 \\
\hline                                  
\end{tabular}
\tablebib{
(K07)~\citet{K07}; (K14) \citet{K14}; (NIST10) \citet{NIST10}.
}
\end{table*}

\section{The RAVE DR5 catalogue of seismic calibrated gravities for giant stars}
\begin{figure}
   \centering
   \includegraphics[width=1\columnwidth]{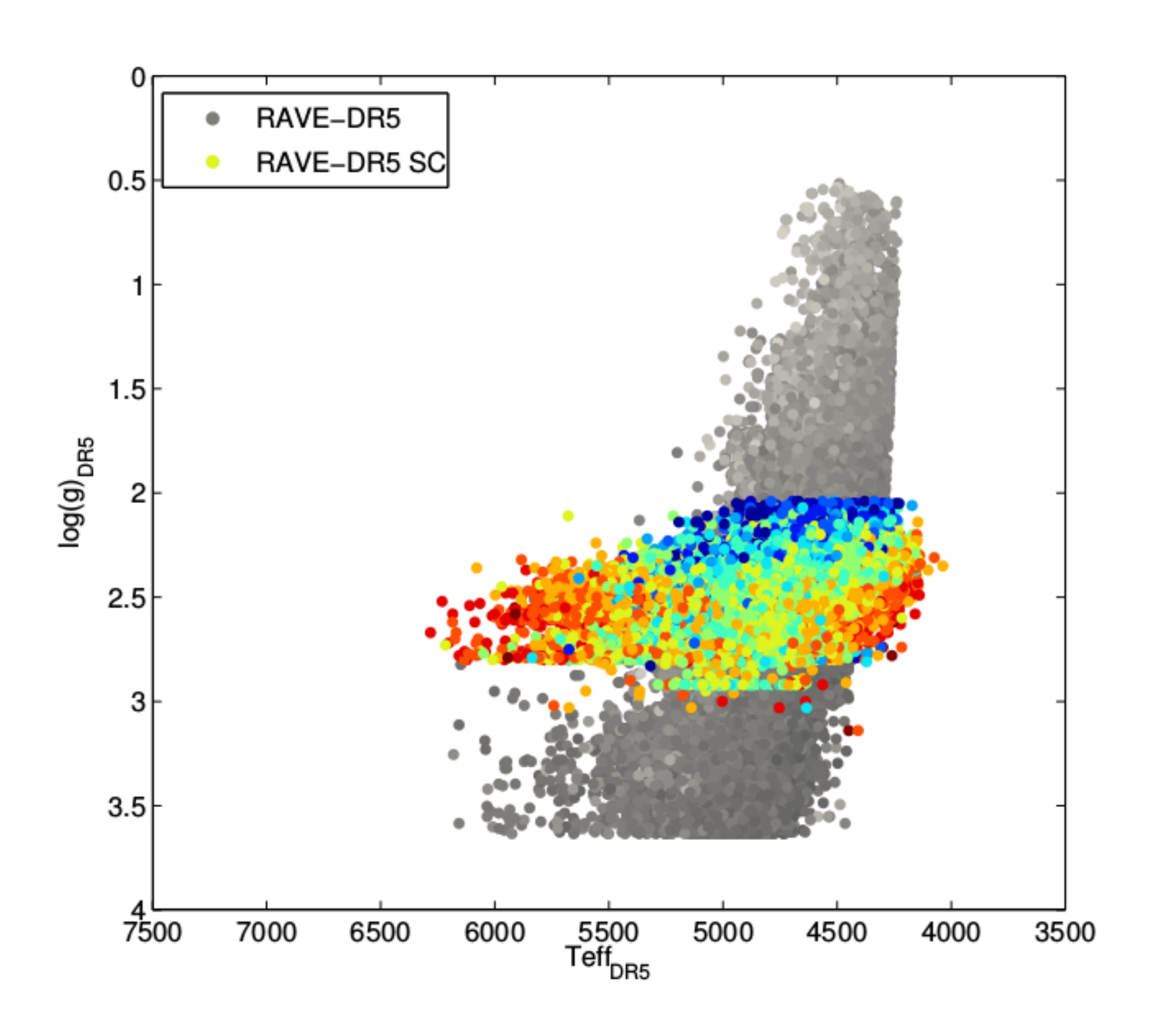}
    \caption{\logg-\teff diagram of the 105,102 seismic calibrated stars in RAVE\_DR5, selected using FLAG\_G$=$1. The diagram is constructed using original DR5 parameters (grey dots) and DR5\_SC parameters (coloured dots). DR5 data is coloured in grey scale, with intensity following metallicity (metal-poor stars are light grey, dark grey marks metal-rich stars). The colour code for DR5\_SC stars follows the standard scale, with metal-poor stars coloured in blue and metal-rich stars coloured in red.}
              \label{Fig:CMD}%
\end{figure}

\end{appendix}

\end{document}